\documentclass[]{article}
\usepackage{amssymb,amsfonts,amsthm,bm,graphicx,amsmath,float,color,longtable}
\restylefloat{figure}
\usepackage[round,numbers,sort&compress]{natbib}
\usepackage[ansinew]{inputenc}

\usepackage{helvet,times}
\usepackage{bm,textcomp}
\usepackage{xr-hyper}
\usepackage{hyperref}

\begin{document}

\title{ATP concentration regulates enzyme kinetics}

\author{Jasmine NIRODY \ and Padmini RANGAMANI \footnote{\textbf {Corresponding author}: padmini.rangamani@berkeley.edu} 
\\[0.05in]
\sl University of California, Berkeley, CA 94720, USA
\rm \normalsize}
\maketitle
\tableofcontents

\begin{abstract} 
{Adenosine 5'-triphosphate (ATP) is the nearly ubiquitous ``energy currency'' of living organisms, and thus is a crucial participant in the majority of enzymatic reactions. The standard models in enzyme kinetics generally ignore the temporal dynamics of ATP because it is assumed to be present in large excess. However, this assumption may not hold in many situations of cellular stress where ATP concentrations may be comparable to substrate levels. Here, we demonstrate the importance of ATP concentration on the dynamics of multi-enzyme reactions by explicit consideration of ATP as a secondary substrate for an enzyme. We apply our model to the mitogen-activated protein (MAP) kinase cascade, which is involved in the regulation of a vast range of cellular activities. We show that three fundamental features of this signaling network --- (i) duration of response, (ii) signal amplification, and (iii) ultrasensitivity to stimulus concentration --- are all dependent on ATP concentration. Our results indicate that the concentration of ATP regulates the response of the MAP kinase activation network, and potentially suggests another possible mechanism for disruption of the cascade in pathogenic states.}
\end{abstract}

\section{Introduction}\label{sec:intro}

Information transfer in cellular systems often occurs through biochemical signaling pathways. A representative scheme of such pathways is one where an external stimulus, often a ligand (e.g., a hormone or another small molecule), binds to a cell surface receptor and activates a series of downstream components (likely associated enzymes). This eventually leads to a change in concentration of a second messenger and subsequently transcription factor activation \citep{neves2002g,kholodenko2008spatio,kholodenko2010signalling}.  

Mathematical models that represent reaction networks as a series of ordinary differential equations have been successfully used to capture the dynamics of such signaling interactions \cite{bhalla1999emergent,neves2002modeling,kholodenko2008spatio}. Perhaps one of the biggest successes in using this approach is that the modeling efforts not only explain the observed experimental time course but also make predictions that can be experimentally tested. These have led to the identification of emergent behavior and motifs in signaling networks \citep{bhalla1999emergent,neves2002modeling,gunawardena2014models}.

One of the most popularly studied signaling networks is the mitogen-activated protein (MAP) kinase cascade.  \citet{huang1996ultrasensitivity} used mass-action kinetics to explain the ultrasensitive response of this cascade to an external stimulus that they observed in their experiments with \textit{Xenopus} oocytes. Mathematical modeling also predicted the existence of bistability in MAP kinase activation in response to the MAPK phosphotase concentration \citep{markevich2004signaling,ortega2006bistability}. This prediction was  experimentally verified \citep{wang2006bistability}. Likewise, the existence of oscillatory dynamics in response to feedback loops was predicted theoretically \citep{kholodenko2000negative}, and later observed experimentally \citep{nakayama2008fgf,shankaran2009rapid}. The MAP kinase cascade is but one prime example of how the power of mathematical modeling can be harnessed to elucidate the properties of biochemical signaling networks \citep{neves2002g,neves2002modeling,kholodenko2010signalling}. Mathematical modeling of reaction networks has thus been the cornerstone of understanding information processing in cellular signaling systems. However, if we are to rely on the predictions of any mathematical model, we must first confirm that we can rely on its assumptions~\cite{gunawardena2014models}.

The standard operating procedure for modeling biochemical signaling networks has been to use mass-action type kinetics to represent binding and stoichiometric reactions and Michaelis-Menten kinetics to describe enzymatic reactions. These approaches have work well, as long as the assumptions underlying these kinetics are satisfied. Mass action kinetics assumes that the concentrations of the reactants are present in large amounts such that the reaction rate is proportional to the concentrations of the reactants. Michaelis-Menten kinetics is derived from mass-action kinetics by enforcing a time scale separation  \citep{gunawardena2014review}. In this study, we relax the assumptions underlying the Michaelis-Menten model and explore the resulting dynamics. 

The Michaelis-Menten model is comprised of the following reaction scheme \cite{michaelis1913kinetik}:
\begin{equation}\label{MMscheme}
\text{E} + \text{S} \rightleftharpoons \text{E-S} \rightarrow \text{E} + \text{P},
\end{equation}
where E is the enzyme, S is the substrate, and P is the product. This model makes two primary assumptions:
\begin{itemize}
\itemsep0em 
\item The conversion of the enzyme-substrate complex to the product and free enzyme is irreversible. 
\item The concentration of ATP is present in large amounts and therefore can be ignored.
\end{itemize}

\noindent The first assumption violates the principle of detailed balance; to overcome this, enzyme kinetics can be modeled using mass-action kinetic schemes  \citep{dill2003molecular}. In this case, the concentration of ATP is usually assumed to be saturating and subsumed within the rate constants \citep{xu2012realistic,gunawardena2014review}. 

Regardless of the choice of kinetic scheme, the above approaches have provided us insight into the relationship between the product concentration and the substrate concentration. However, an open question remains: what happens to enzyme kinetics under different ATP concentrations? In many states of cellular stress, including metabolic distress and apoptosis \citep{jennings1985nucleotide,richter1996control,leist1997intracellular,schutt2012moderately}, ATP concentration can be limiting. Experiments have shown that the concentration of ATP can affect the dynamics of several cellular processes, including caspase activation \citep{verrax2011intracellular}, the insulin response \citep{kim2008role}, and the movement of endocytic vesicles along microtubules \citep{murray2000reconstitution}. Such examples suggest that modeling efforts might need to consider the effect of ATP concentration on dynamics of signaling networks. 

In this work, we study the role of ATP concentration in regulating enzyme kinetics. We model enzymes as two-substrate catalysts --- the first is the substrate itself, and the second is ATP --- to examine the effect of explicit inclusion of ATP in regulating enzyme kinetics. We explore the effects of ATP concentration in simple signaling modules and compare each module outcome with the corresponding Michaelis-Menten response. We then apply our approach to the MAP kinase signaling network and show how the concentration of ATP can influence several significant properties of this kinase cascade.
\begin{figure}[t]
\centering
\includegraphics[width=0.8\textwidth]{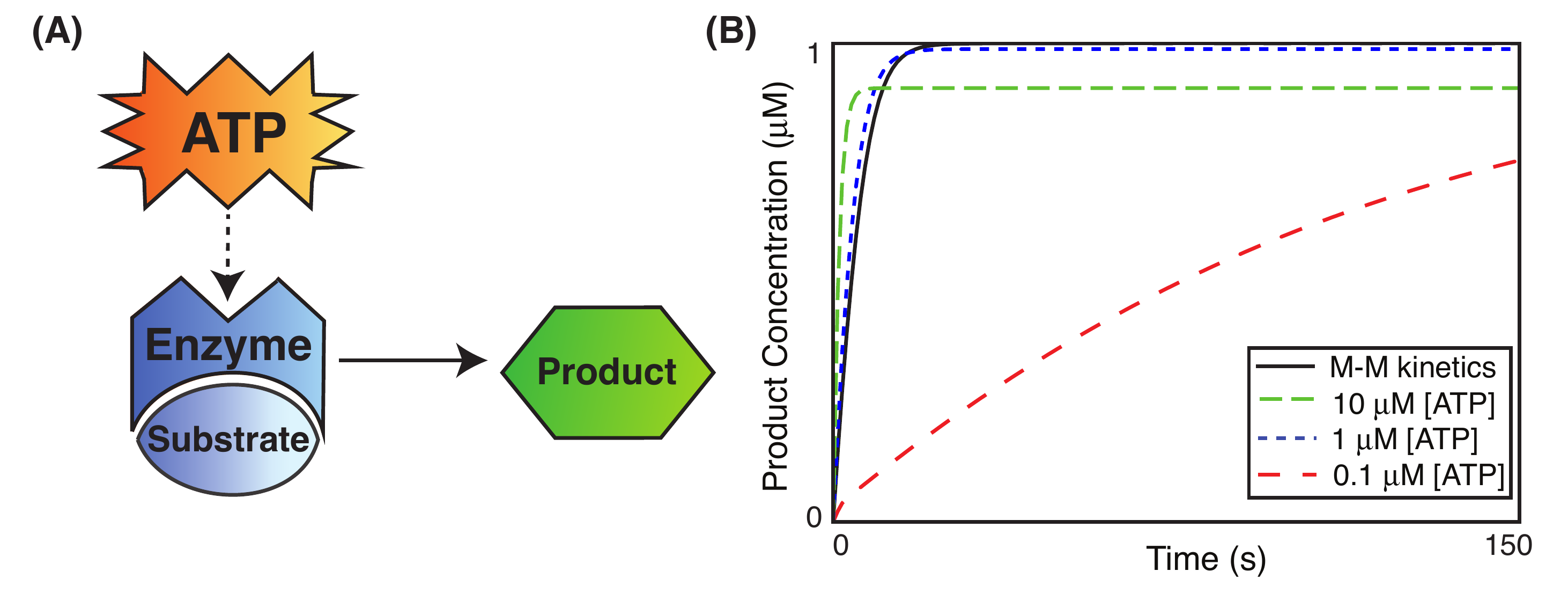}
\caption{\textit{(A)} An enzyme catalyzed reaction that converts the substrate into a product also needs ATP. How does ATP concentration affect enzyme kinetics? \textit{(B)} Comparison of Michaelis-Menten kinetics with our model for single-enzyme catalyzed reaction. Mid-range values of [ATP] result in dynamics very similar to those predicted by the Michaelis-Menten model. Dynamics for high and low ATP concentrations diverge from Michaelis-Menten kinetics due to sequestration of enzyme in complex form and depletion of ATP, respectively.}
\label{fig:one_enzyme}
\end{figure}

\section{Model Development and Implementation}
\subsection{Overview and single-enzyme reactions}
For a single enzyme catalyzing the conversion of a substrate to a product, the Michaelis-Menten model presumes the simple two-step reaction shown in Equation \ref{MMscheme}. A physiological requirement for the application of Michaelis-Menten kinetics is that the formation and disassociation of the enzyme-substrate complex E-S is rapid \citep{gunawardena2012some}, allowing for the quasi-steady-state assumption $\frac{d[\text{E-S}]}{dt} \approxeq 0$. This results in the following reaction flux from the substrate to the product:
\begin{equation}
- \frac{d[\text{S}]}{dt} = \frac{V_{\text{max}} [\text{S}]}{K_m+[\text{S}]} = \frac{d[\text{P}]}{dt}.
\end{equation}
Here, [S] and [P] are the concentrations (in $\mu$M) of the substrate and product, respectively; $K_m$ is the Michaelis constant (with units $\mu$M); and $V_{\text{max}}$ is the maximal flux of the reaction (with units $\mu$M s$^{-1}$).
\begin{figure}
\centering
\includegraphics[width=0.75\textwidth]{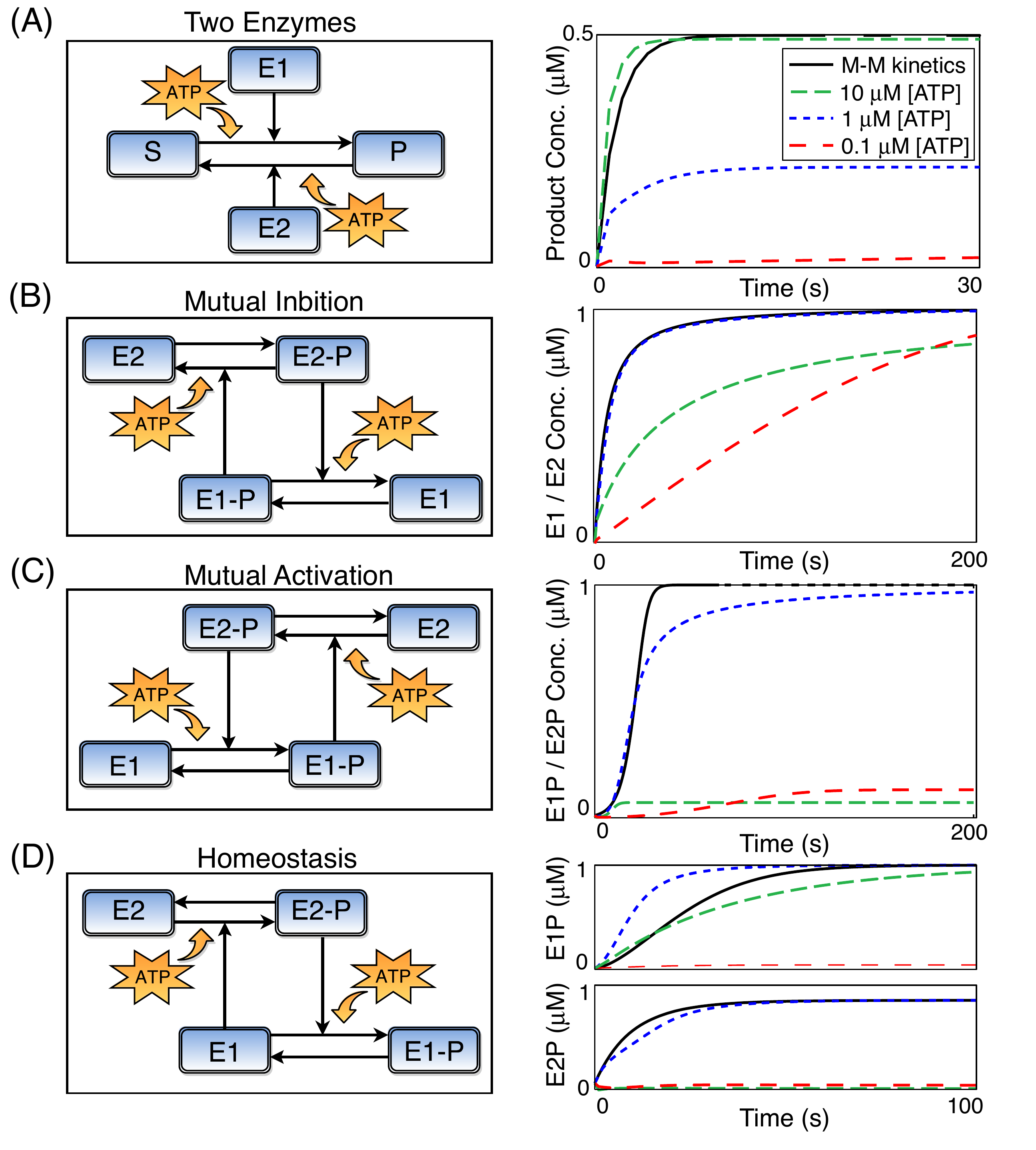}
\caption{Comparison of Michaelis-Menten kinetics with our model for (A) two separate enzymes catalyzing the forward and backward reaction, (B) Mutual inhibition loop, (C) Mutual activation loop, and (D) Homeostasis. Depending on the module interactions, the system can be highly sensitive to ATP concentration or insensitive to ATP concentration.}
\label{fig:loops}
\end{figure}

For a single-enzyme catalyzed reaction, we consider ATP explicitly as a second substrate (Figure~\ref{fig:one_enzyme}A). To account for both parallel and sequential binding schemes, we allow for both reactions to take place; the full reaction scheme considered is provided in Table \ref{table:oneenzyme}. The resulting transition complex, E-ATP-S, subsequently releases the product and ADP, thus regenerating the enzyme. Rather than include the details of various metabolic pathways, ATP regeneration is modeled as a single mass-action reaction where ADP and inorganic phosphate combine to form ATP.  

\subsection{Two-enzyme signaling modules}
Enzymes do not act in isolation. Therefore, in addition to the single-enzyme catalyzed reaction, we consider several two-enzyme modules that appear frequently in biological signaling networks. Figure \ref{fig:loops}A depicts a two-enzyme reaction in which one enzyme catalyzes the forward reaction (S $\rightarrow$ P) and the other catalyzes the backward reaction (P $\rightarrow$ S). The schemes depicted in Figure \ref{fig:loops}B-D are modified from \citet{tyson2003sniffers}.

Figures \ref{fig:loops}B and \ref{fig:loops}C depict two possible forms of positive feedback. In the first, \textit{mutual inhibition} (Figure \ref{fig:loops}B), E$_1$ and E$_2$ are mutually antagonistic. That is, each species facilitates the degradation of the species that facilitates its own degradation. The second, \textit{mutual activation} (Figure \ref{fig:loops}C), is the more traditionally recognized form of a positive feedback loop: each species helps generate the species that aids its generation. An example of negative feedback is \textit{homeostasis} (Figure \ref{fig:loops}D) where a species (E$_2$P) facilitates the degradation of the enzyme (E$_1$) responsible for its generation. Full reaction schemes used for the above modules are shown in Tables~\ref{table:twoenzyme} - \ref{table:homeo}.

\subsection{MAP kinase activation network}

The mitogen-activated protein kinase (MAP kinase) cascade is a signaling cascade consisting of three kinases: MAPK kinase kinase (MAPKKK, KKK, or Raf), MAPK kinase (MAPKK, KK, or MEK), and MAPK (K or Erk) . A stimulus (either a small GTP-binding protein such as Ras or another kinase such as PKC) activates MAPKKK to initiate the cascade. Singly phosphorylated MAPKKK  (MAPKKK-P or MAPKKK*) activates MAPKK by phosphorylation at two serine residues, and doubly phosphorylated MAPKK (MAPKK-PP or MAPKK**) activates MAPK by phosphorylation at threonine and tyrosine residues \citep{nishida1993map,huang1996ultrasensitivity,chang2001mammalian,roskoski2012erk1}.

Though its basic structure is highly conserved, the MAPK signaling cascade is associated with a large variety of biological responses. Consequently, this network has spurred the development of several mathematical models, the most well-known of which was developed by \citet{huang1996ultrasensitivity} to describe MAPK activation in \textit{Xenopus} oocytes. \citet{bhalla1999emergent} also considered the MAPK network in their study of second messenger cascades in neurons. Their model contains a negative feedback loop in addition to the Huang and Ferrell's proposed model: MAPK** inhibits the cascade by doubly phosphorylating MAPKKK*.

\begin{figure}
\centering
\includegraphics[width=0.75\textwidth]{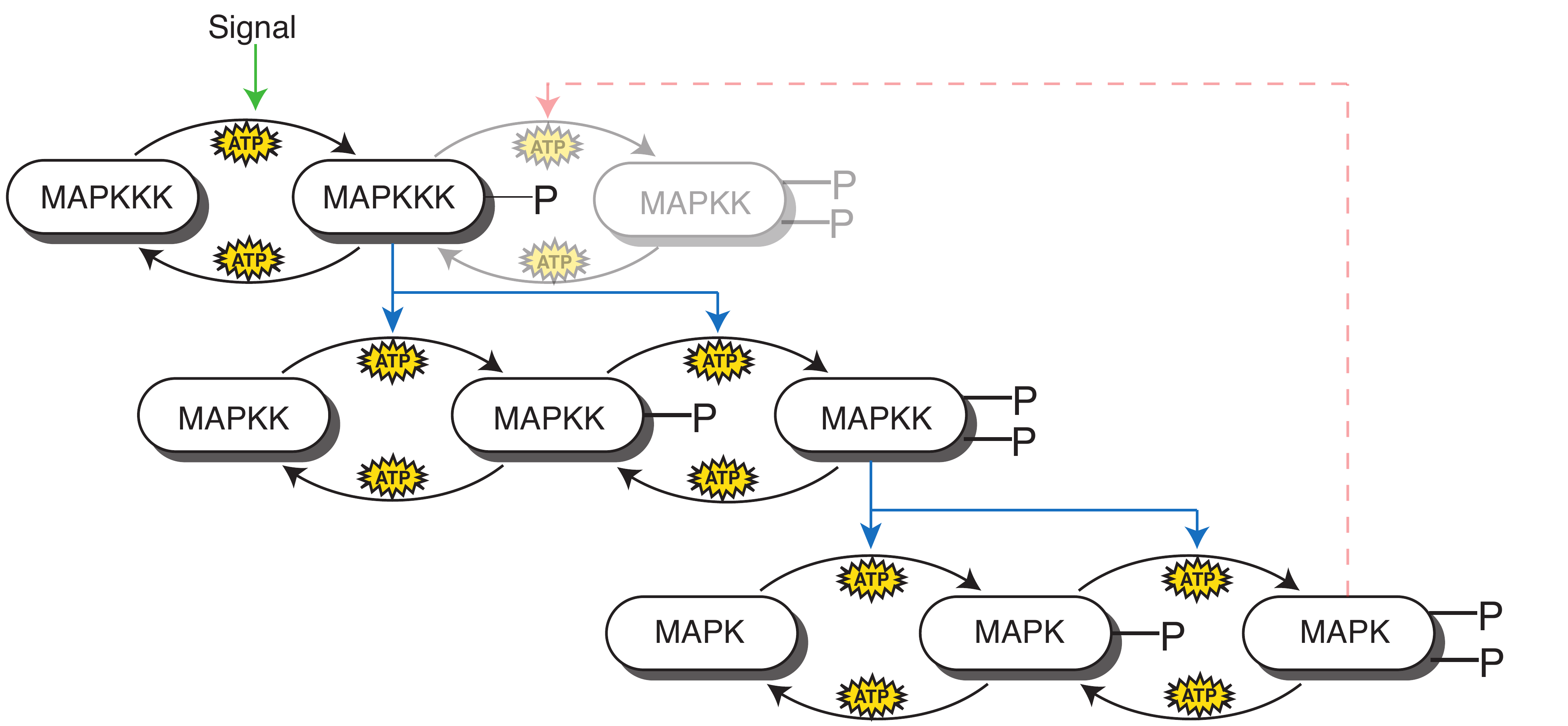}
\caption{MAP kinase activation network showing the different phosphorylation and dephosphorylation steps. The main model is adapted from \citet{huang1996ultrasensitivity}. The red dashed arrow and the grayed out module of MAPK-PP are the feedback loop used in \citet{bhalla1999emergent}.}
\label{fig:Mapk}
\end{figure}

Figure \ref{fig:Mapk} provides an overview of the various phosphorylation and dephosphorylation events involved in the network. The feedback loop described by Bhalla and Iyengar is shown in light gray. Initial concentrations and kinetic parameters are taken from \citet{bhalla1999emergent}. As their model assumed Michaelis-Menten kinetics, we chose mass-action kinetic parameters ($k_{\text{on}}$ and $k_{\text{off}}$) for each module of the network which best matched the corresponding Michaelis constants while maintaining realistic physiological responses for the overall cascade. Reaction schemes and kinetic parameters used for the base model by \citet{huang1996ultrasensitivity} are shown in Table~\ref{table:mapk}; Table \ref{table:mapkfb} describes the additional feedback module from \citet{bhalla1999emergent}.

\subsection{Implementation}

All reactions were modeled using mass action kinetics and the resulting ordinary differential equations were solved using the stiff solvers `ode15s' and `ode23s' routines in MATLAB$^{\text{\textregistered}}$ (Mathworks, Natick, MA). For comparison, we also constructed and simulated Michaelis-Menten kinetics for the single-enzyme reaction in Figure \ref{fig:one_enzyme} and for all the modules shown in Figure~\ref{fig:loops}.

For all the signaling modules, we set rates such that all ``forward'' reactions are significantly faster than the associated reverse reactions. To further destabilize the transition complex, we multiply the forward reaction rate $k_\text{on}$ for its disassociation in all modules by $10^3$. Kinetic parameters used for each module are provided in the corresponding Supplementary tables. All MATLAB code will be made available on request. Kinetic parameters for the MAP kinase reaction network were chosen as described above \citep{bhalla1999emergent}. 

\section{Effect of ATP concentration on independent signaling modules}

In the following sections, we explore the effect of ATP concentration on independent motifs which are commonly found nested within large signaling networks. In particular, we characterize how limiting energy in the form of low ATP concentration can lead to non-Michaelis-Menten kinetics under various network topologies (e.g., systems with positive or negative feedback).

\subsection{Reversible reactions catalyzed by one or two enzymes}
We first consider a reversible reaction involving catalysis by either a single enzyme or by two enzymes. In the single-enzyme case, the enzyme acts only in the forward direction (S $\rightarrow$ P), but all reactions follow the law of mass action, and are reversible. This is in contrast to the Michaelis-Menten model, where the reverse reaction rate for the scheme E-S $\rightarrow$ P + E is set to 0. This is clear from the time course for Michaelis-Menten kinetics (Figure \ref{fig:one_enzyme}B, solid black line): the product is formed until all substrate is converted.

When ATP is explicitly considered, we find that an ATP concentration on the same order of magnitude as the initial substrate concentration (the timecourse for [ATP]$|_{t=0}$ = [S]$|_{t=0}$ =  1 $\mu$M is shown by a blue line in Figure \ref{fig:one_enzyme}B) matches Michaelis-Menten kinetics quite well (though the equilibrium value falls short of the Michaelis-Menten equilibrium value of 1 $\mu$M). When ATP is limiting ([ATP]$|_{t=0}$ = 0.1 $\mu$M, shown by the red line in Figure \ref{fig:one_enzyme}) the reaction proceeds far more slowly and takes significantly longer to reach equilibrium. When ATP is available in excess ([ATP]$|_{t=0}$ = 10 $\mu$M, shown by the green line in Figure \ref{fig:one_enzyme}), the equilibrium value of formed product is lower than that observed for mid-range ATP concentrations, though the time to reach equilibrium is shorter.

The monotonic decrease in time to equilibrium with increase in ATP concentration is an expected feature of mass action kinetics: a high ATP concentration leads to a rapid formation of the transition complex, and consequently to a faster formation of product. The lower equilibrium value of product in environments with excess ATP compared to the Michaelis-Menten equilibrium is due to the inclusion of a reverse reaction for the complex dissociation (see Tables \ref{table:oneenzyme} and \ref{table:twoenzyme}). While the Michaelis-Menten kinetic scheme faithfully reproduces the dynamics of many equilibrated chemical systems, it does not uphold the principle of detailed balance: the formation of product is not balanced by any reverse reaction \citep{dill2003molecular}. Removing this irreversibility assumption results in a non-zero concentration of substrate complexes at equilibrium. Furthermore, as the sequestration of enzyme in complexes (e.g., the transition complex or E-ADP) is ignored by Michaelis-Menten kinetics, a slight decrease in production past some ``optimal'' ATP concentration is also observed in the full mass-action system.

We further consider an extension of the same reversible reaction, where both the forward (S $\rightarrow$ P) and backward (P~$\rightarrow$~S) reactions are catalyzed by two distinct enzymes, E$_1$ and E$_2$. Given symmetry of the forward and backward reactions, Michaelis-Menten kinetics result in equal equilibrium concentrations of substrate and product in this system (the time course for the generation of product is shown by the black line in Figure \ref{fig:loops}A). When ATP is explicitly considered, we observed that the time to equilibrium decreases with increasing ATP levels, as in the single-enzyme model.

\begin{figure}
\centering
\includegraphics[width=0.8\textwidth]{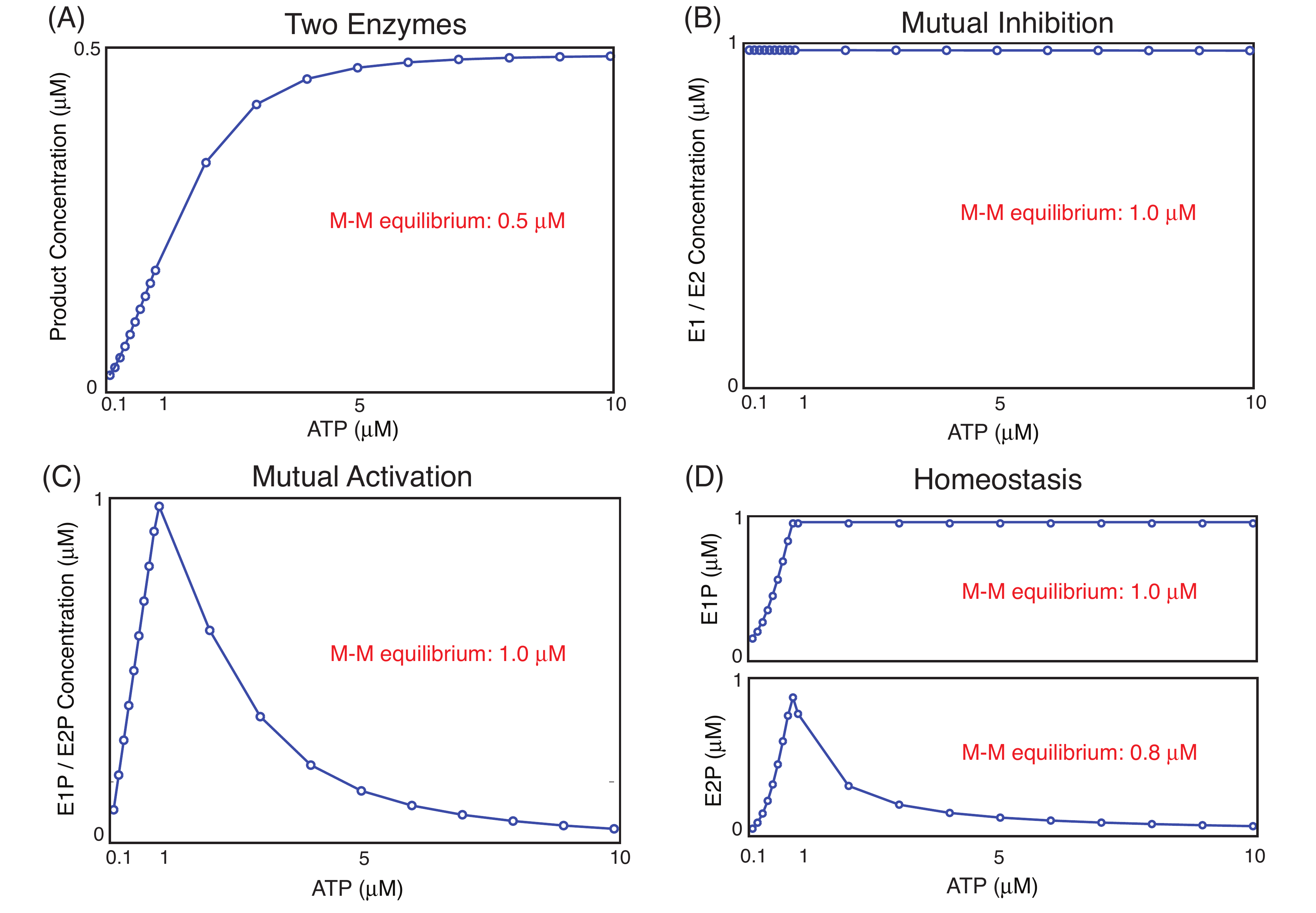}
\caption{Equilibrium concentrations of product or responses from the different module structure in response to different concentration of ATP. The equilibrium values for the Michaelis-Menten model with equivalent initial conditions are provided for comparison. \textit{(A)} The concentration of product is sensitive to the initial concentration of ATP at low ATP concentrations in the two enzyme model. \textit{(B)} The response of the mutual inhibition module, as measured by E$_1$P and E$_2$P concentration is independent of the ATP concentration. \textit{(C)} The mutual activation module and (D) the homeostasis modules are sensitive to ATP concentrations and have an optimal response at an intermediate concentration of ATP.}
\label{fig:ATP_concn}
\end{figure}

Similarly to the single-enzyme case, Michaelis-Menten kinetics always result in higher concentrations at equilibrium of the products of both reactions (here, S and P) relative to the full mass action system. However, in contrast to the single-enzyme case, our chosen ``medium'' level of ATP (equal to that of the initial substrate concentration, [ATP]$|_{t=0}$ = [S]$|_{t=0}$ =~1~$\mu$M) is not sufficient to elicit the full response of the system. Figure \ref{fig:ATP_concn} shows the equilibrium concentration of the ``product'' of each module as a function of ATP concentration; simulations were run for 20,000 s to ensure equilibration. As shown in Figure~\ref{fig:ATP_concn}A, the system response saturates only at an ATP concentration of approximately 8 $\mu$M. As ATP levels rise even further, the system response decreases, albeit slightly, just as in the single-enzyme system (see Figure \ref{sfig:highATP2enz} in the Supplementary Material). As seen previously, due to complex formation and lack of turnover, even the maximum response of the reversible system is lower than that of the Michaelis-Menten model (Figure \ref{fig:ATP_concn}). This over-prediction of system response by Michaelis-Menten kinetics due to sequestration has been observed experimentally as well~\citep{frank2010out}.

\subsection{Mutual inhibition and mutual activation}
Figures \ref{fig:loops}B and \ref{fig:loops}C show two examples of simple positive feedback modules involving the phosphorylation and dephosphorylation of two enzymes. In Figure \ref{fig:loops}B (\textit{mutual inhibition}), the phosphorylated version of each enzyme catalyzes the dephosphorylation of the other---each species depletes the species that depletes it. Figure \ref{fig:loops}C (\textit{mutual activation}) is an example of more ``traditional'' positive feedback: each phosphorylated enzyme stimulates the phosphorylation of the other---each species generates the species that generates it.

For both systems, the dynamics associated with a medium ATP concentration ([ATP]$|_{t=0}$ = 1 $\mu$M) are most in line with that of the Michaelis-Menten model. As seen previously, Michaelis-Menten kinetics overestimate the conversion to product in reversible systems. However, we find that the equilibrium value of dephosphorylated enzyme---the ``product''---is independent of ATP concentration (Figure \ref{fig:ATP_concn}B) for mutual inhibition. In contrast, the equilibrium concentration of product of mutual activation (phosphorylated enzyme) appears dependent on the concentration of ATP (Figure \ref{fig:ATP_concn}C). Note that as the two reactions within each module are symmetric, the dynamics associated with both enzymes are equivalent. 

This discrepancy arises from two differences between the two schemes. Firstly, the mutual inhibition motif is comprised of two dephosphorylation reactions, while mutual activation is comprised of two phosphorylations. The regular release of inorganic phosphate $P_i$ during the time course of mutual inhibition results in rapid recycling of ATP. When ATP concentration is low, the lack of recycling in the mutual activation module results in fast depletion of ATP (see Figure \ref{sfig:atp} in the Supplementary Material). Secondly, the products of the forward (i.e., enzyme-catalyzed) reactions in the mutual inhibition module are the dephosphorylated enzymes E$_1$ and E$_2$, which do not form complexes with ATP. In contrast, the forward reactions in the ``mutual activation'' module are phosphorylations, which result in the production of the catalysts E$_1$P and E$_2$P. These products are promptly sequestered in complexes with ATP and ADP (see Figure \ref{sfig:complex} in the Supplementary Material). A study on the molecular chaperone GroEL under saturating ATP conditions found that the rate of ADP release is important in determining the rate of generation and equilibrium concentration of product \citep{frank2010out}.

However, while the equilibrium concentration of the product may be largely independent of ATP concentration, the convergence time to equilibrium is not: ATP concentrations higher and lower than 1 $\mu$M result in a slower convergence to equilibrium (Figure~\ref{fig:loops}B). With low initial ATP concentrations, cycling assures that there is a constant (albeit shallow) energy pool for reactions to continue. However, the maintained concentration of ATP is too low for a speedy convergence to equilibrium. When ATP concentration is very high, a signifiant portion of the enzyme very quickly becomes isolated in complexes, particularly E-ADP (Figure \ref{sfig:complex}), and the slow breakdown of these complexes contributes to a slower convergence to equilibrium. Though the lower, ``optimal'' concentrations of ATP also result in the formation of such complexes, the initial amounts formed are significantly lower than with excess ATP, allowing for the presence of more free enzyme to continue catalysis.

\subsection{Homeostasis}
In contrast to the first two signaling modules considered, Figure \ref{fig:loops}D depicts a case of negative feedback (\textit{homeostasis}): enzyme E$_2$P phosphorlyates (and thus deactivates) E$_1$, the catalyst for its creation. In the right panel of the figure, we show time courses for both E$_1$P and E$_2$P; because, unlike the other modules, the two reactions are not symmetric. Because it suppresses its own production, Michaelis-Menten kinetics predict slower dynamics and a lower equilibrium value for E$_2$P as compared to other modules (Figure \ref{fig:loops}D).

The equilibrium concentrations for E$_2$P for higher and lower ATP concentrations are significantly lower than that observed for mid-range ATP levels. This is once again due to isolation of enzyme in complex form (for high initial ATP) and rapid energy depletion (for low initial ATP). Interestingly, however, mid-range energy levels, result in \textit{higher} equilibrium concentrations of E$_2$P than predicted by the Michaelis-Menten model (Figure \ref{fig:ATP_concn}D). 

The saturation of E$_2$P is due to the conversion of E$_1$ to E$_1$P, as is shown in the time course of the reactions (Figure \ref{fig:loops}D). In the reversible model, however, a (small) subset of E$_1$ sequestered in complex form continues to facilitate the conversion of E$_2$ to E$_2$P. This sequestration and subsequent slow release of free enzyme results in a long period of low influx into a small but non-negligible pool of catalytic units. Since such accumulation of complex is neglected in Michaelis-Menten kinetics, the E$_2$P equilibrium concentration predicted by assuming irreversibility is actually lower than in the full reversible model at certain ATP concentrations. This suggests that, while in most cases, sequestration of enzyme simply results in equilibrium concentrations that are lower than expected, it may result in other indirect effects in certain network topologies \cite{roussel1998slowly}.

\begin{figure}[!!h]
\centering
\includegraphics[width=0.8\textwidth]{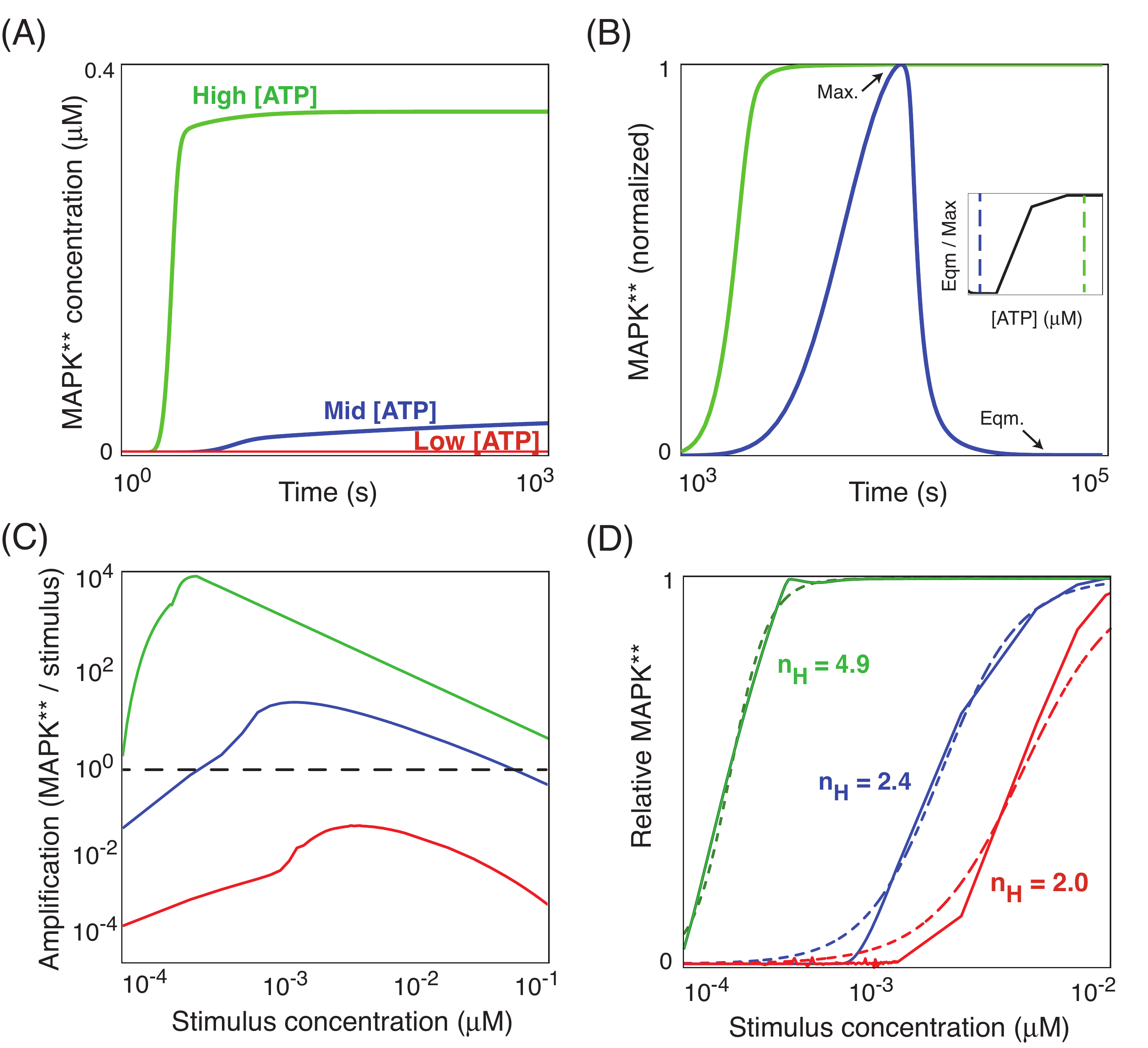}
\caption{Effect of ATP concentration on the MAP kinase activation network by \citet{huang1996ultrasensitivity}. High ([ATP] = 0.6~$\mu$M), medium ([ATP] = 0.2 $\mu$M), and low ([ATP] = 0.05 $\mu$M) ATP concentrations are shown throughout in green, blue, and red, respectively. \textit{(A)} Time courses of MAPK activation given a strong stimulus (E$_1 = 0.1$ $\mu$M) in high, medium, and low ATP environments. \textit{(B)} Effect of ATP concentration on duration of MAPK activation. Time courses of MAPK activation given a relatively weak stimulus (E$_1 = 10^{-4}$ $\mu$M) are shown for high and medium ATP levels. Because of the disparity in amplitude of responses between ATP concentrations considered, MAPK** levels are normalized to be between 0 and 1. Inlay shows the transition from transient to sustained response as a function of [ATP]. Green and blue dashed lines denote the concentrations used to generate the time courses shown. \textit{(C)} Effect of ATP concentration on the signal amplification by the MAPK cascade. The ordinate axis shows the ratio of maximum activated MAPK** to input stimulus; the stimulus concentration $(\mu$M) is given on the abscissa. The dashed black line marks 1 (no amplification). \textit{(D)} Effect of ATP on ultrasensitivity in the MAPK network. Solid lines show predicted steady-state responses for various ATP levels; corresponding dashed lines show Hill curves with comparable steepness. Hill coefficients ($n_H$) are reported for each curve.}
\label{fig:MAPKres1}
\end{figure}

\section{Effect of ATP Concentration on the MAP Kinase Cascade}
Biological signaling networks are composed of many modules. It is often the interplay between these modules within a highly organized structure that is responsible for the observed characteristics of a system. A prime example of this is the MAP kinase cascade, a signaling network that has been extensively studied both experimentally and theoretically because of its unique emergent properties. In their seminal work, \citet{huang1996ultrasensitivity} identified the ultrasensitivity in this cascade. Subsequently, this network has been shown to have bistability in the presence of a feedback loop \citep{bhalla1999emergent,kholodenko2000negative}. While there have been extensive mathematical models of epidermal growth factor (EGF) and other agonist-induced stimulation of the MAP kinase cascade, the role of ATP in facilitating the cascade has not been explicitly considered. In the following, we are concerned with how the concentration of ATP affects the signaling characteristics of MAP kinase activation. Specifically, we study three fundamental properties of this ubiquitous network: (i) time course and duration of response, (ii) ultrasensitivity, and (iii) signal amplification \citep{mayawala2004mapk}.

\subsection{Time course and duration of MAPK activation}

Since there are multiple sequential steps involved in the MAP kinase cascade, the time between the application of signal and the appearance of activated MAPK is not negligible. Figure \ref{fig:MAPKres1}A shows time courses for activated MAPK (i.e., MAPK**) after the application of a strong stimulus ([E$_1]|_{t=0}$ = 0.1 $\mu$M). The amplitude of the response increases with ATP concentration, while the time to maximal response decreases as ATP levels rise. Additionally, the steepness of the response increases with ATP concentration; at the highest ATP concentration considered, the activation of MAPK is more switch-like than in lower ATP environments.

Several mechanisms for dynamic temporal regulation have been suggested to act on the MAP kinase cascade; for an overview, we refer the reader to several excellent reviews on the topic~\citep{neves2002g,hao2007systems,kholodenko2008spatio,kholodenko2010signalling}. Here, we considered how regulation by the negative feedback loop proposed by \citet{bhalla1999emergent} is affected by ATP concentration. In this feedback system, activated MAPK doubly phosphorylates MAPKKK*, stripping it of its catalytic abilities and down-regulating the cascade. The components of this regulatory loop are shown in faded grey in Figure \ref{fig:Mapk}.

When the feedback is applied at the strength proposed by \citet{bhalla1999emergent}, it is effective only when ATP concentration is high; at lower concentrations, little discernible difference in the time course was observed (see Figure \ref{sfig:MAPKfdbk} in the Supplementary Material). At a lower signal strength, we also found that strengthened feedback resulted in a transient activation of MAPK, as opposed to a sustained response when feedback was not present (Figure \ref{sfig:MAPKfdbk2}).

The transition between transient and sustained activation has been shown to occur in a switch-like manner with respect to signal strength \citep{hao2007systems}. However, transient responses were also obtained without feedback for stronger, persistent signals when ATP concentration was low (see Figures \ref{fig:MAPKres1}B and \ref{sfig:MAPKfdbk}C). Furthermore, we observed that the shift from a transient to sustained response is sharp with respect to ATP concentration as well (see inlay of Figure \ref{fig:MAPKres1}B). This may point to another method to achieve specificity in MAPK pathways \citep{sabbagh2001specificity}. This transition in particular has been famously implicated in cell fate decisions: transient activation of MAPK by EGF leads to cell proliferation, while sustained activation in response to nerve growth factor leads to cell differentiation \citep{marshall1995specificity}. 
\subsection{Amplification}
As mentioned above, a common objective of signaling cascades is stimulus amplification. Figure \ref{fig:MAPKres1}C shows the system amplification (calculated as the ratio of the concentrations of MAPK** to E$_1$) as a function of stimulus concentration. Maximal amplification in the Michaelis-Menten model (apparent Hill coefficient, $n_H = 1$) is achieved in the case of a low saturated kinase, while in more sigmoidal responses ($n_H > 1$) maximal amplification occurs near K$_{10}$, the stimulus concentration generating 10\% of the maximum response \citep{bluthgen2001map}. The peak amplificiation is found approximately at this point for all three ATP concentrations considered ([ATP]$|_{t=0} = 0.6$ $\mu$M, 0.2 $\mu$M, and 0.05 $\mu$M; shown in green, blue, and red in Figure \ref{fig:MAPKres1}C, respectively).

In principle, a three-tiered cascade can generate an immense degree of signal amplification; that is, a very small concentration of signal (here E$_1$) can bring about a far larger concentration of activated MAPK** \citep{koshland1982amplification,mayawala2004mapk}. Indeed, when ATP concentration is sufficiently high (green line, Figure \ref{fig:MAPKres1}C), we see that the MAPK** concentration can be over 10,000 times that of the input stimulus.

In practice, however, such high amplification is not necessary and is rarely observed---most experimental studies find approximately a 10- to 30-fold response relative to stimulus \citep{fujioka2006dynamics,roskoski2012erk1,leng2004effects,moxham1996jun}. Under high ATP concentrations, this range of amplification occurs when the stimulus concentration is relatively high compared to the concentration of kinase, which is often the case in experiments \citep{huang1996ultrasensitivity,fujioka2006dynamics}. However, at mid-range ATP concentrations (blue line, Figure \ref{fig:MAPKres1}C; [ATP]$|_{t=0} = 0.2$ $\mu$M $\sim$ [kinase]$|_{t=0}$), the amplification observed in physiological systems is very close to the maximum that can be obtained, leaving far less room for error in stimulus concentration. Furthermore, in an ATP-limited environment (red), the cascade fails to amplify any input, regardless of signal strength.
\subsection{Ultrasensitivity}
While the most intuitive function of the MAP kinase cascade is signal amplification, its crucial involvement in cellular decision-making processes such as cell cycle control and cell fate induction points to another fundamental feature: the generation of a ultrasensitive switch-like response to stimuli \citep{olson1995essential,bluthgen2001map}. In \textit{Xenopus} oocytes, MAP kinase cascade indeed acts as an all-or-none switch. \citet{huang1996ultrasensitivity} predicted that MAP kinase activation was best predicted by a Hill coefficient of approximately 5, a robust result which holds for a large range of kinetic parameters. However, both their model calculations and experiments were performed with ATP at constant concentration (i.e., abundantly available such that depletion could be considered negligible).

At high ATP concentrations, our simulations confirm their results ([ATP]$|_{t=0}$ = 0.6 $\mu$M, shown in green in Figure \ref{fig:MAPKres1}D), producing a stimulus-response curve comparable to a Hill curve with $n_H = 4.9$. Apparent Hill coefficients were calculated as 
\begin{equation}
n_H = \frac{\log(81)}{\log(\text{K}_{90} / \text{K}_{10})},
\end{equation}
where K$_{90}$ and K$_{10}$ denote the stimulus concentrations that generate 90\% and 10\% of the maximum system response, respectively \citep{bluthgen2001map}. Note that both K$_{90}$ and K$_{10}$, as well as their ratio, vary with ATP concentration (Figure~\ref{fig:MAPKres1}C). 
However, at lower ATP concentrations ([ATP]$|_{t=0}$ = 0.2 $\mu$M and 0.05 $\mu$M, respectively shown in blue and red in Figure~\ref{fig:MAPKres1}D), we observed a significant decrease in the apparent Hill coefficient, indicating a steady decay of the all-or-none response to a more graded one. Zero-order ultrasensitivity has been found to be partially responsible for the switch-like response of the MAP kinase network \citep{huang1996ultrasensitivity,ferrell1996tripping}. This mechanism is observed only when the enzyme is operating near saturation---that is, when [S] $>>$ [E] \citep{goldbeter1981amplified}. Because ATP acts as a secondary substrate, our observed decline of the all-or-none response in the MAP kinase cascade with ATP concentration may be attributed to the loss of the contribution from zero-order ultrasensitivity. 

\section{Discussion}
Organisms respond to their environment by performing computations through signaling networks in response to mechanical or chemical stimuli. Due to the incredible complexity of these computations, mathematical models of reaction networks often employ certain simplifications in order to generate interpretable results. One such model, the Michaelis-Menten scheme, has stood the test of time to emerge as the standard model in enzyme kinetics---many scientific articles provide values for the Michaelis constant $K_m$, with little or no mention of the traditional kinetic constants from elementary mass-action~\citep{michaelis1913kinetik}. 

However, while Michaelis-Menten kinetics quite faithfully reproduce the equilibrium dynamics of many enzymatic processes, we must be acutely aware of the assumptions used to derive it. In particular, the Michaelis-Menten model relies crucially on time-scale separation: we must assume that the formation of enzyme-substrate complex and the subsequent release of product and free enzyme occur so rapidly that the amount of sequestered enzyme is negligible \citep{gunawardena2014review}. Furthermore, the standard Michaelis-Menten model assumes that secondary substrates (e.g., ATP) are available in excess so as not to be rate-limiting; their dynamics are generally ignored along with those of the enzyme. 

The removal of this ``single-substrate enzyme'' approximation, however, immediately calls into question the near-universal application of Michaelis-Menten kinetics: if ATP (or any other secondary substrate) is required for formation of the transition complex E-ATP-S and subsequent release of product, then it must play a crucial role in enzyme kinetics. In particular, ATP is not always abundant in the cellular milieu \citep{taylor1986energetics,leist1997intracellular,kim2008role}, which can lead to the sequestration of a non-negligible amount of enzyme in E-S complex form. Such sequestration, unsurprisingly, results in a time course that diverges significantly from that predicted by Michaelis-Menten kinetics \citep{bluthgen2006effects}. For instance, lack of ATP cycling results in the isolation of enzyme in E-ADP complexes and a consequent decrease in the amount of activated target from predicted values~\citep{frank2010out}.

At this point, it is good to note an important assumption made in \textit{our} model. Throughout this paper, we have utilized the principle of mass-action. That is, we have assumed that the dynamics considered involve concentrations that are adequately characterized by large molecular particle numbers. Assuming a local volume of $\sim$1 $\mu$L, even a concentration of 10$^{-7}\mu$M (far lower than any we have considered in this paper) corresponds to approximately 60,000 molecules. However, inclusion of randomness is a natural step towards understanding any biological system. In particular, an analysis of the effect of ATP concentration on the noise-filtering ability of an enzyme cascade would be interesting given our results on the deterioration of ultrasensitivity in the MAP kinase network with decreasing ATP levels \citep{gonze2002robustness,bluthgen2003robust} and will be explored in a future work.

Despite the immense complexity of most signaling networks, several simple motifs and pathways often serve as biochemical ``building blocks'', and are conserved and used across molecular functions and cell types. In this paper, we have examined the role of ATP concentration in modulating the dynamics of enzymatic processes under some of these common network topologies. We surprisingly find that, while limiting ATP concentrations usually result in a lowered or delayed system response, certain network structures (e.g., the negative feedback module shown in Figure \ref{fig:loops}D) may be able to exploit a small range of ``medium'' ATP concentrations to \textit{increase} their output relative to the system response in ATP-rich environments. 

Such bidirectional regulation by ATP concentration, where both low and high ATP levels result in reduced function, has been observed in several cellular processes, including dynein walking \citep{hirakawa2000processive} and the ubiquitin-proteasome system \citep{huang2010physiological}. In the latter, the basal level of ATP was shown to constitutively repress proteasome function, allowing for the upregulation of the proteasome at ATP concentrations characteristic in cellular stress ($\sim$10-100 $\mu$M). This upregulation has also been implicated in pathogenesis. For instance, muscle wasting observed in acute respiratory distress syndrome (ARDS) is linked to increased proteasome activity \citep{epstein1996mechanisms,ottenheijm2006activation}. Furthermore, this regulatory role of ATP suggests that raising intracellular ATP concentration may inhibit proteasome activity in cancer cells. This could serve as an alternative treatment to pharmacological inhibitors, which often have negative side effects \citep{orciuolo2007unexpected}. Our results point to one possible mechanism underlying the bidirectional control of ATP concentration on protein activity. 

We further explored the effect of ATP concentration on the MAP kinase cascade, a signaling module which is activated in response to almost any extra- or intracellular stimulus. Our results suggest that the ATP concentration modulates at least three fundamental properties of the cascade: (i) duration of the response, (ii) signal amplification, and (iii) ultrasensitivity with respect to stimulus concentration. The duration and amplitude of activation have been shown computationally to be sensitive to changes in the phosphatase reactions at the MAPKK and MAPK level, respectively \citep{mayawala2004mapk}. As ATP is involved in reactions at all levels, it is not surprising that these properties are dependent on its concentration. 

In contrast, ultrasensitivity has been previously shown to be robust to a wide range of kinetic parameters under the assumption of constant ATP concentration \citep{huang1996ultrasensitivity}. We found, however, that this switch-like response to stimulus concentration fades into a more graded one as ATP concentration decreases. As ATP is a substrate of the reaction, enzymes operate at saturation only when the ATP concentration is high (a prerequisite for zero-order ultrasensitivity). Furthermore, as our model considers reversible mass-action kinetics, the sequestration of enzyme and substrate in complex form results in a further loss of zero-order ultrasensitivity \citep{markevich2004signaling,bluthgen2006effects}. Our results suggest that ATP concentration may serve as another possible mechanism for MAPK regulation in normal cells, or for disruption of the cascade in pathological states. 

Thus far, we have assumed that all enzymes have equivalent access to ATP. Signaling pathways are highly compartmentalized, with certain receptors bound to membrane and others freely floating in the cytosol. For instance in the MAPK cascade, kinases are membrane-bound, associated with scaffolding proteins, while phosphatases are cytosolic \citep{dhanasekaran2007scaffold}. As the individual steps of the cascade have been associated independently with certain observed system properties \citep{mayawala2004mapk}, the regulatory effects of heterogenous ATP distribution are certainly worth exploring.  

\section*{ACKNOWLEDGMENTS}
We would like to thank the American Institute of Mathematics for organizing and inviting us to participate in their 2013 workshop, \textit{``Mathematical problems arising from biochemical reaction networks".} The idea of exploring the role of ATP concentration in enzyme reactions arose from discussions during this workshop. We also acknowledge funding from an NSF IGERT from the CiBER center at UC Berkeley (to J.A.N.) and the University of California Chancellor's Postdoctoral Fellowship (to P.R.).

\newpage

\setcounter{table}{0}
        \renewcommand{\thetable}{S\arabic{table}}%
        \setcounter{figure}{0}
        \renewcommand{\thefigure}{S\arabic{figure}}%

\section*{Supplementary Material}
\subsection*{Tables}
For all modules, the units for $k_{\text{on}}$ and $k_{\text{off}}$ must be such that $v = k_{\text{on}}\sum_i{R_i^{m_i}} - k_{\text{off}}\sum_i{P_i^{n_i}}$ (where $v$ has units $\mu$M s$^{-1}$) is balanced. Here, $R_i$ and $P_i$ denote reactant and product species and $m_i$ and $n_i$ denote their respective balanced coefficients. For example, if a reaction has two reactants and one product species, $k_{\text{on}}$ has units $\mu$M$^{-1}$ s$^{-1}$ and $k_{\text{off}}$ has units s$^{-1}$. Likewise, if a reaction has one reactant and two (three) products, 
$k_{\text{on}}$ has units s$^{-1}$ and $k_{\text{off}}$ has units $\mu$M$^{-1}$ s$^{-1}$ ($\mu$M$^{-2}$ s$^{-1}$).\\

\noindent All initial concentrations that are not provided are set to $0$.

\begin{center}
\begin{table*}[!!h]
\centering
\caption{Reactions for one enzyme two substrate model}
\begin{tabular} {l l l l }
\hline
\multicolumn{4}{l}{Initial concentrations (all in $\mu$M): [E] = 1; [S] = 1; [P] = 0; [Ph] = [ATP]}\\
\hline
Reaction &  k$_{\text{on}}$ & k$_{\text{off}}$ & Notes\\ [0.5ex]
\hline
E + S $\rightleftharpoons$ E-S & 1.0 & 0.2 & Enzyme binds substrate\\
E + ATP $\rightleftharpoons$ E-ATP & 1.0  & 0.2 & Enzyme binds ATP\\
E-ATP + S $ \rightleftharpoons$ E-ATP-S & 1000 & 0.2 & Enzyme-ATP complex binds substrate\\
E-S + ATP $\rightleftharpoons$ E-ATP-S & 1.0 & 0.2 & Enzyme-substrate complex binds ATP\\
E-ATP-S $\rightleftharpoons$ E-ADP + P + $\text{P}_{i}$ & 1.0 & 0.2 & Catalysis; product is released\\
E-ADP $\rightleftharpoons$ E + ADP & 1.0 & 0.2 & Disassociation of ADP from enzyme\\
E-ATP $\rightleftharpoons$ E-ADP + $\text{P}_{i}$ & 1.0 & 0.2 & Autocatalysis\\
ATP $\rightleftharpoons$ ADP + $\text{P}_{i}$ & 1.0 & 0.2 & Regeneration of ATP (metabolism)\\
\hline
\multicolumn{4}{l}{Michaelis-Menten parameters: $V_{\text{max}} = 0.1$ $\mu\text{M s}^{-1}$; $K_m = 1.0$ $\mu\text{M}$}\\
\hline
\end{tabular}
\label{table:oneenzyme}
\end{table*}

\begin{table*}[h]
\centering
\caption{Reactions for two enzyme (reversible) model}
\begin{tabular} {l l l l }
\hline
\multicolumn{4}{l}{Initial concentrations (all in $\mu$M): [E$_1$] = [E$_2$] = 1; [S] = 1; [P] = 0; [Ph] = [ATP]}\\
\hline
Reaction &  k$_{\text{on}}$ & k$_{\text{off}}$ & Notes\\ [0.5ex]
\hline
E$_1$ + S $\rightleftharpoons$ E$_1$-S & 0.001 & 1e-4 & Enzyme 1 binds substrate\\
E$_1$  + ATP $\rightleftharpoons$ E$_1$-ATP & 0.001 & 1e-4 & Enzyme 1 binds ATP\\
E$_1$-ATP + S $ \rightleftharpoons$ E$_1$-ATP-S & 0.001 & 1e-4 & Enzyme-ATP complex binds substrate\\
E$_1$-S + ATP $\rightleftharpoons$ E$_1$-ATP-S & 0.001 & 1e-4 & Enzyme-substrate complex binds ATP\\
E$_1$-ATP-S $\rightleftharpoons$ E$_1$-ADP + P + $\text{P}_{i}$ & 1.0 & 1e-4 & Catalysis; product is released\\
E$_1$-ADP $\rightleftharpoons$ E$_1$  + ADP & 0.001 & 1e-4 & Disassociation of ADP from Enzyme 1\\
E$_1$-ATP $\rightleftharpoons$ E$_1$-ADP + $\text{P}_{i}$ & 0.001 & 1e-4 & Autocatalysis\\
E$_2$ + P $\rightleftharpoons$ E$_2$-P & 0.001 & 1e-4 & Enzyme 2 binds product of first reaction\\
E$_2$  + ATP $\rightleftharpoons$ E$_2$-ATP & 0.001  & 1e-4 & Enzyme 2 binds ATP\\
E$_2$-ATP + P $ \rightleftharpoons$ E$_2$-ATP-P & 0.001 & 1e-4 & Enzyme-ATP complex binds ``product''\\
E$_2$-P + ATP $\rightleftharpoons$ E$_2$-ATP-P & 0.001 & 1e-4 & Enzyme-``product'' complex binds ATP\\
E$_2$-ATP-P $\rightleftharpoons$ E$_2$-ADP + S + $\text{P}_{i}$ & 1.0 & 1e-4 & Catalysis; original substrate is released\\
ATP $\rightleftharpoons$ ADP + $\text{P}_{i}$ & 0.001 & 1e-4 & Regeneration of ATP (metabolism)\\
\hline
\multicolumn{4}{l}{Michaelis-Menten parameters: $V_{\text{max}} = 0.1$ $\mu\text{M s}^{-1}$; $K_m = 1.0$ $\mu\text{M}$}\\
\hline
\end{tabular}
\label{table:twoenzyme}
\end{table*}

\begin{table*}[h!]
\centering
\caption{Reactions for mutual inhibition model \citep{tyson2003sniffers}}
\begin{tabular} {l l l l }
\hline
\multicolumn{4}{l}{Initial concentrations (all in $\mu$M): [E$_1$P]  = [E$_2$P] = 0.99; [E$_1$]  = [E$_2$] = 0.01; [Ph] = [ATP]}\\
\hline
Reaction &  k$_{\text{on}}$ & k$_{\text{off}}$ & Notes\\ [0.5ex]
\hline
E$_1$P + E$_2$P $\rightleftharpoons$ E$_1$P-E$_2$P & 1.0 & 1e-4 & Enzyme 1 binds substrate\\
E$_1$P  + ATP $\rightleftharpoons$ E$_1$P-ATP & 1.0 & 1e-4 & Enzyme 1 binds ATP\\
E$_1$P-ATP + E$_2$P $ \rightleftharpoons$ E$_1$P-ATP-E$_2$P & 1.0 & 1e-4 & Enzyme-ATP complex binds substrate\\
E$_1$P-E$_2$P + ATP $\rightleftharpoons$ E$_1$P-ATP-E$_2$P & 1.0 & 1e-4 & Enzyme 1-substrate complex binds ATP\\
E$_1$P-ATP-E$_2$P $\rightleftharpoons$ E$_1$P-ADP + E$_2$ + 2$\text{P}_{i}$ & 1000 & 1e-4 & Catalysis; product is released\\
E$_1$P-ADP $\rightleftharpoons$ E$_1$P  + ADP & 1.0 & 1e-4 & Disassociation of ADP from Enzyme 1\\
E$_1$P-ATP $\rightleftharpoons$ E$_1$P-ADP + $\text{P}_{i}$ & 1.0 & 1e-4 & Autocatalysis\\
E$_2$P + E$_1$P $\rightleftharpoons$ E$_2$P-E$_1$P & 1.0 & 1e-4 & Enzyme 2 binds substrate\\
E$_2$P  + ATP $\rightleftharpoons$ E$_2$P-ATP & 1.0  & 1e-4 & Enzyme 2 binds ATP\\
E$_2$P-ATP + E$_1$P $ \rightleftharpoons$ E$_2$P-ATP-E$_1$P & 1.0 & 1e-4 & Enzyme2-ATP complex binds substrate\\
E$_2$-P + ATP $\rightleftharpoons$ E$_2$P-ATP & 1.0 & 1e-4 & Enzyme-substrate complex binds ATP\\
E$_2$-ATP-E$_1$P $\rightleftharpoons$ E$_2$-ADP + E$_1$ + 2$\text{P}_{i}$ & 1000 & 1e-4 & Catalysis; product is released\\
ATP $\rightleftharpoons$ ADP + $\text{P}_{i}$ & 1.0 & 1e-4 & Regeneration of ATP (metabolism)\\
\hline
\multicolumn{4}{l}{Michaelis-Menten parameters: $V_{\text{max}} = 0.3$ $\mu\text{M s}^{-1}$; $K_m = 1.0$ $\mu\text{M}$}\\
\hline
\end{tabular}
\label{table:mutualinhib}
\end{table*}

\begin{table*}[h!]
\centering
\caption{Reactions for mutual activation model \citep{tyson2003sniffers}}
\begin{tabular} {l l l l }
\hline
\multicolumn{4}{l}{Initial concentrations (all in $\mu$M): [E$_1$P]  = [E$_2$P] = 0.01; [E$_1$]  = [E$_2$] = 0.99; [Ph] = [ATP]}\\
\hline
Reaction &  k$_{\text{on}}$ & k$_{\text{off}}$ & Notes\\ [0.5ex]
\hline
E$_1$P + E$_2$ $\rightleftharpoons$ E$_1$P-E$_2$ & 0.001 & 1e-4 & Enzyme 1 binds substrate\\
E$_1$P + ATP $\rightleftharpoons$ E$_1$P-ATP & 0.001 & 1e-4 & Enzyme 1 binds ATP\\
E$_1$P-ATP + E$_2$ $ \rightleftharpoons$ E$_1$P-ATP-E$_2$ & 0.001 & 1e-4 & Enzyme-ATP complex binds substrate\\
E$_1$P-E$_2$ + ATP $\rightleftharpoons$ E$_1$P-ATP-E$_2$ & 0.001 & 1e-4 & Enzyme 1-substrate complex binds ATP\\
E$_1$P-ATP-E$_2$ $\rightleftharpoons$ E$_1$P-ADP + E$_2$P & 1.0 & 1e-4 & Catalysis; product is released\\
E$_1$P-ADP $\rightleftharpoons$ E$_1$P  + ADP & 0.001 & 1e-4 & Disassociation of ADP from Enzyme 1\\
E$_1$P-ATP $\rightleftharpoons$ E$_1$P-ADP + $\text{P}_{i}$ & 0.001 & 1e-4 & Autocatalysis\\
E$_2$P + E$_1$ $\rightleftharpoons$ E$_2$P-E$_1$ & 0.001 & 1e-4 & Enzyme 2 binds substrate\\
E$_2$P  + ATP $\rightleftharpoons$ E$_2$P-ATP & 0.001  & 1e-4 & Enzyme 2 binds ATP\\
E$_2$P-ATP + E$_1$ $ \rightleftharpoons$ E$_2$P-ATP-E$_1$ & 0.001 & 1e-4 & Enzyme2-ATP complex binds substrate\\
E$_2$P-E$_1$ + ATP $\rightleftharpoons$ E$_2$P-ATP-E$_1$ & 0.001 & 1e-4 & Enzyme-substrate complex binds ATP\\
E$_2$-ATP-E$_1$ $\rightleftharpoons$ E$_2$-ADP + E$_1$P & 1.0 & 1e-4 & Catalysis; product is released\\
ATP $\rightleftharpoons$ ADP + $\text{P}_{i}$ & 0.001 & 1e-4 & Regeneration of ATP (metabolism)\\
\hline
\multicolumn{4}{l}{Michaelis-Menten parameters: $V_{\text{max}} = 0.4$ $\mu\text{M s}^{-1}$; $K_m = 1.0$  $\mu\text{M}$}\\
\hline
\end{tabular}
\label{table:mutualact}
\end{table*}
\onecolumn
\begin{table*}[h!]
\centering
\caption{Reactions for homeostasis model \citep{tyson2003sniffers}}
\begin{tabular} {l l l l }
\hline
\multicolumn{4}{l}{Initial concentrations (all in $\mu$M): [E$_1$]  = [E$_2$] = 0.9; [E$_1$P]  = [E$_2$P] = 0.1; [Ph] = [ATP]}\\
\hline
Reaction &  k$_{\text{on}}$ & k$_{\text{off}}$ & Notes\\ [0.5ex]
\hline
E$_1$ + E$_2$ $\rightleftharpoons$ E$_1$-E$_2$ & 1.0 & 1e-4 & Enzyme 1 binds substrate\\
E$_1$ + ATP $\rightleftharpoons$ E$_1$-ATP & 1.0 & 1e-4 & Enzyme 1 binds ATP\\
E$_1$-ATP + E$_2$ $ \rightleftharpoons$ E$_1$-ATP-E$_2$ & 1.0 & 1e-4 & Enzyme-ATP complex binds substrate\\
E$_1$-E$_2$ + ATP $\rightleftharpoons$ E$_1$-ATP-E$_2$ & 1.0 & 1e-4 & Enzyme 1-substrate complex binds ATP\\
E$_1$-ATP-E$_2$ $\rightleftharpoons$ E$_1$-ADP + E$_2$P & 1000 & 1e-4 & Catalysis; product is released\\
E$_1$-ADP $\rightleftharpoons$ E$_1$  + ADP & 1.0 & 1e-4 & Disassociation of ADP from Enzyme 1\\
E$_1$-ATP $\rightleftharpoons$ E$_1$-ADP + $\text{P}_{i}$ & 1.0 & 1e-4 & Autocatalysis\\
E$_2$P + E$_1$P $\rightleftharpoons$ E$_2$P-E$_1$P & 1.0 & 1e-4 & Enzyme 2 binds substrate\\
E$_2$P  + ATP $\rightleftharpoons$ E$_2$P-ATP & 1.0  & 1e-4 & Enzyme 2 binds ATP\\
E$_2$P-ATP + E$_1$ $ \rightleftharpoons$ E$_2$P-ATP-E$_1$P & 1.0 & 1e-4 & Enzyme2-ATP complex binds substrate\\
E$_2$-P + ATP $\rightleftharpoons$ E$_2$P-ATP & 1.0 & 1e-4 & Enzyme-substrate complex binds ATP\\
E$_2$-ATP-E$_1$P $\rightleftharpoons$ E$_2$-ADP + E$_1$ + $\text{P}_{i}$ & 1000 & 1e-4 & Catalysis; product is released\\
ATP $\rightleftharpoons$ ADP + $\text{P}_{i}$ & 1.0 & 1e-4 & Regeneration of ATP (metabolism)\\
\hline
\multicolumn{4}{l}{Michaelis-Menten parameters: $V_{\text{max}} = 0.1$ $\mu\text{M s}^{-1}$; $K_m = 1.0$ $\mu\text{M}$}\\
\hline
\end{tabular}
\label{table:homeo}
\end{table*}

\begin{longtable}{|l|l|l|l|}
\caption {Reactions for MAPK Model \citep{huang1996ultrasensitivity}}
\label{table:mapk} \vspace{-0.3cm} \\ \hline
\multicolumn{4}{|l|}{Initial concentrations (all in $\mu$M): [E$_2$]  = 0.224; [E$_3$] = 0.0032; [KKK]  = 0.2; [KK] = 0.18; [K] = 0.36; [Ph] = [ATP]}\\ 
\hline \multicolumn{1}{|c|}{\textbf{Reaction}} & \multicolumn{1}{c|}{\textbf{k$_{\text{on}}$}} & \multicolumn{1}{c|}{\textbf{k$_{\text{off}}$}} & \multicolumn{1}{c|}{\textbf{Notes}}\\ \hline 

\endfirsthead

\multicolumn{4}{c}%
{{\bfseries \tablename\ \thetable{} -- continued from previous page}} \\
\hline \multicolumn{1}{|c|}{\textbf{Reaction}} & \multicolumn{1}{c|}{\textbf{k$_{on}$}} & \multicolumn{1}{c|}{\textbf{k$_{off}$}} & \multicolumn{1}{c|}{\textbf{Notes}} \\ \hline 
\endhead

\hline \multicolumn{4}{|r|}{{Continued on next page}} \\ \hline
\endfoot

\hline
\endlastfoot
\hline
\multicolumn{4}{|c|}{\bf Metabolism} \\
\hline
ATP $\rightleftharpoons$ ADP + $\text{P}_{i}$ & 1.0 & 0.001 & Regeneration of ATP (metabolism)\\
\hline
\multicolumn{4}{|c|}{\bf KKK phosphorylation} \\
\hline
E$_1$+ KKK $\rightleftharpoons$ E$_1$-KKK & 0.04 & 0.001 & Enzyme binds substrate\\
E$_1$  + ATP $\rightleftharpoons$ E$_1$-ATP & 0.04  & 0.001 & Enzyme binds ATP\\
E$_1$-ATP + KKK $ \rightleftharpoons$ E$_1$-ATP-KKK & 0.04 & 0.001 & Enzyme-ATP complex binds substrate\\
E$_1$-KKK + ATP $\rightleftharpoons$ E$_1$-ATP-KKK & 0.04 & 0.001 & Enzyme-substrate complex binds ATP\\
E$_1$-ATP-KKK $\rightleftharpoons$ KKK-P + E$_1$-ADP & 0.04 & 1e-30 & KKK is phosphorlyated\\
E$_1$-ADP $\rightleftharpoons$ E$_1$ + ADP & 0.04 & 1e-30 & Enzyme is regenerated\\
\hline
\multicolumn{4}{|c|}{\bf KKK-P dephosphorylation}\\
\hline
E$_2$ + KKK-P $\rightleftharpoons$ E$_2$-KKK-P & 0.1 & 0.001 & Enzyme binds substrate\\
E$_2$  + ATP $\rightleftharpoons$ E$_2$-ATP & 0.1  & 0.001 & Enzyme binds ATP\\
E$_2$-ATP + KKK-P $ \rightleftharpoons$ E$_2$-ATP-KKK-P & 0.1 & 0.001 & Enzyme-ATP complex binds substrate\\
E$_2$-KKK-P + ATP $\rightleftharpoons$ E$_2$-ATP-KKK-P & 0.1 & 0.001 & Enzyme-substrate complex binds ATP\\
E$_2$-ATP-KKK-P$_i$ $\rightleftharpoons$ KKK + E$_2$-ADP + 2P$_i$ & 0.1 & 1e-30 & KKK-P is dephosphorlyated\\
E$_2$-ADP $\rightleftharpoons$ E$_2$ + ADP & 0.1 & 1e-30 & Enzyme is regenerated\\
\hline
\multicolumn{4}{|c|}{\bf KK mono-phosphorylation}\\
\hline
KKK-P+ KK$\rightleftharpoons$ KKK-P-KK & 1.0 & 0.001 & Enzyme binds substrate\\
KKK-P + ATP $\rightleftharpoons$ KKK-P-ATP & 1.0  & 0.001 & Enzyme binds ATP\\
KKK-P-ATP + KK $ \rightleftharpoons$ KKK-P-ATP-KK & 1.0 & 0.001 & Enzyme-ATP complex binds substrate\\
KKK-P-KK + ATP $\rightleftharpoons$ KKK-P-ATP-KK & 1.0 & 0.001 & Enzyme-substrate complex binds ATP\\
KKK-P-ATP-KK $\rightleftharpoons$ KK-P + KKK-P-ADP & 1.0 & 1e-30 & KK is phosphorlyated\\
KKK-P-ADP $\rightleftharpoons$ KKK-P + ADP & 1.0 & 1e-30 & Enzyme is regenerated\\
\hline
\multicolumn{4}{|c|}{\bf KK-P dephosphorylation}\\
\hline
E$_2$+ KK-P $\rightleftharpoons$ E$_2$-KK-P & 0.01 & 0.001 & Enzyme binds substrate\\
E$_2$-ATP + KK-P $ \rightleftharpoons$ E$_2$-ATP-KK-P & 0.01 & 0.001 & Enzyme-ATP complex binds substrate\\
E$_2$-KK-P + ATP $\rightleftharpoons$ E$_2$-ATP-KK-P & 0.01 & 0.001 & Enzyme-substrate complex binds ATP\\
E$_2$-ATP-KK-P $\rightleftharpoons$ KK + E$_2$-ADP + 2P$_i$ & 0.01 & 1e-30 & KK-P is dephosphorylated\\
\hline
\multicolumn{4}{|c|}{\bf KK-P (double) phosphorylation}\\
\hline
KKK-P+ KK-P $\rightleftharpoons$ KKK-P-KK-P & 1.0 & 0.001 & Enzyme binds substrate\\
KKK-P-ATP + KK-P $ \rightleftharpoons$ KKK-P-ATP-KK-P & 1.0 & 0.001 & Enzyme-ATP complex binds substrate\\
KKK-P-KK-P + ATP $\rightleftharpoons$ KKK-P-ATP-KK-P & 1.0 & 0.001 & Enzyme-substrate complex binds ATP\\
KKK-P-ATP-KK-P $\rightleftharpoons$ KK-PP + KKK-P-ADP & 1.0 & 1e-30 & KK-P is phosphorlyated\\
\hline
\multicolumn{4}{|c|}{\bf KK-PP (mono) dephosphorylation}\\
\hline
E$_2$+ KK-PP $\rightleftharpoons$ E$_2$-KK-PP & 1.0 & 0.1 & Enzyme binds substrate\\
E$_2$-ATP + KK-PP $ \rightleftharpoons$ E$_2$-ATP-KK-PP & 1.0 & 0.1 & Enzyme-ATP complex binds substrate\\
E$_2$-KK-PP + ATP $\rightleftharpoons$ E$_2$-ATP-KK-PP & 1.0 & 0.1 & Enzyme-substrate complex binds ATP\\
E$_2$-ATP-KK-PP $\rightleftharpoons$ KK-P + E$_2$-ADP + 2P$_i$ & 1.0 & 1e-30 & KK-PP is dephosphorlyated\\
\hline
\multicolumn{4}{|c|}{\bf K mono-phosphorylation}\\
\hline
KK-PP+ K$\rightleftharpoons$ KK-PP-K & 5.0 & 0.001 & Enzyme binds substrate\\
KK-PP+ ATP $\rightleftharpoons$ KK-PP-ATP & 5.0  & 0.001 & Enzyme binds ATP\\
KK-PP-ATP + K $ \rightleftharpoons$ KK-PP-ATP-K & 5.0 & 0.001 & Enzyme-ATP complex binds substrate\\
KK-PP-K + ATP $\rightleftharpoons$ KK-PP-ATP-K & 5.0 & 0.001 & Enzyme-substrate complex binds ATP\\
KK-(P$_i$)$_2$-ATP-K $\rightleftharpoons$ K-P +KK-PP-ADP & 5.0 & 1e-30 & K is phosphorlyated\\
KK-P-ADP $\rightleftharpoons$ KK-P + ADP & 5.0 & 1e-30 & Enzyme is regenerated\\
\hline
\multicolumn{4}{|c|}{\bf K-P dephosphorylation}\\
\hline
E$_3$+ K-P $\rightleftharpoons$ E$_3$-K-P& 1.0 & 0.001 & Enzyme binds substrate\\
E$_3$  + ATP $\rightleftharpoons$ E$_3$-ATP & 1.0  & 0.001 & Enzyme binds ATP\\
E$_3$-ATP + K-P $ \rightleftharpoons$ E$_3$-ATP-K-P & 1.0 & 0.001 & Enzyme-ATP complex binds substrate\\
E$_3$-K-P + ATP $\rightleftharpoons$ E$_3$-ATP-K-P & 1.0 & 0.001 & Enzyme-substrate complex binds ATP\\
E$_3$-ATP-K-P $\rightleftharpoons$ K + E$_3$-ADP + 2P$_i$ & 1.0 & 1e-30 & K-P is dephosphorlyated\\
E$_3$-ADP $\rightleftharpoons$ E$_3$ + ADP & 1.0 & 1e-30 & Enzyme is regenerated\\
\hline
\multicolumn{4}{|c|}{\bf K-P (double) phosphorylation}\\
\hline
KK-PP+ K-P $\rightleftharpoons$ KK-PP-K-P & 5.0 & 0.001 & Enzyme binds substrate\\
KK-PP-ATP + K-P $ \rightleftharpoons$KK-PP-ATP-K-P & 5.0 & 0.001 & Enzyme-ATP complex binds substrate\\
KK-PP-K-P + ATP $\rightleftharpoons$ KK-PP-ATP-K-P & 5.0 & 0.001 & Enzyme-substrate complex binds ATP\\
KK-PP-ATP-K-P $\rightleftharpoons$ K-PP +KK-PP-ADP & 4.0 & 0.1 & K is phosphorlyated\\
KKK-P-ADP $\rightleftharpoons$ KKK-P + ADP & 1.0 & 1e-30 & Enzyme is regenerated\\
\hline
\multicolumn{4}{|c|}{\bf K-PP (mono) dephosphorylation}\\
\hline
E$_3$+ K-PP $\rightleftharpoons$ E$_3$-K-PP & 1.0 & 0.001 & Enzyme binds substrate\\
E$_3$-ATP + K-PP $ \rightleftharpoons$ E$_3$-ATP-K-PP & 1.0 & 0.001 & Enzyme-ATP complex binds substrate\\
E$_3$-K-PP + ATP $\rightleftharpoons$ E$_3$-ATP-K-PP & 1.0 & 0.001 & Enzyme-substrate complex binds ATP\\
E$_3$-ATP-K-PP $\rightleftharpoons$ K-P + E$_3$-ADP + 2P$_i$ & 1.0 & 1e-30 & K-PP is dephosphorlyated\\
\end{longtable}

\newpage

\begin{longtable}{|l|l|l|l|}
\caption {Feedback loop in MAPK network \citep{bhalla1999emergent}}
\label{table:mapkfb} \vspace{-0.3cm} \\
\hline
\multicolumn{1}{|c|}{\textbf{Reaction}} & \multicolumn{1}{c|}{\textbf{k$_{\text{on}}$}} & \multicolumn{1}{c|}{\textbf{k$_{\text{off}}$}} & \multicolumn{1}{c|}{\textbf{Notes}}\\ \hline 
\endfirsthead
\multicolumn{4}{c}%
{{\bfseries \tablename\ \thetable{} -- continued from previous page}} \\
\hline \multicolumn{1}{|c|}{\textbf{Reaction}} & \multicolumn{1}{c|}{\textbf{k$_{on}$}} & \multicolumn{1}{c|}{\textbf{k$_{off}$}} & \multicolumn{1}{c|}{\textbf{Notes}} \\ \hline 
\endhead
\hline \multicolumn{4}{|r|}{{Continued on next page}} \\ \hline
\endfoot
 \hline
\endlastfoot
\multicolumn{4}{|c|}{\bf KKK-P (double) phosphorylation} \\
\hline
K-PP + KKK-P $\rightleftharpoons$ K-PP-KKK-P & 0.1 & 0.001 & Enzyme binds substrate\\
K-PP  + ATP $\rightleftharpoons$ K-PP-ATP & 0.1  & 0.001 & Enzyme binds ATP\\
K-PP-ATP + KKK-P $ \rightleftharpoons$ K-PP-ATP-KKK & 0.1 & 0.001 & Enzyme-ATP complex binds substrate\\
K-PP-KKK-P + ATP $\rightleftharpoons$ K-PP-ATP-KKK-P & 0.1 & 0.001 & Enzyme-substrate complex binds ATP\\
K-PP-ATP-KKK-P $\rightleftharpoons$ KKK-PP + K-PP-ADP & 0.1 & 1e-30 & KKK-P is phosphorlyated\\
K-PP-ADP $\rightleftharpoons$ K-PP + ADP & 0.1 & 1e-30 & Enzyme is regenerated\\
\hline
\multicolumn{4}{|c|}{\bf KKK-PP dephosphorylation}\\
\hline
E$_2$ + KKK-PP $\rightleftharpoons$ E$_2$-KKK-PP & 0.1 & 0.001 & Enzyme binds substrate\\
E$_2$  + ATP $\rightleftharpoons$ E$_2$-ATP & 0.1  & 0.001 & Enzyme binds ATP\\
E$_2$-ATP + KKK-PP $ \rightleftharpoons$ E$_2$-ATP-KKK-PP & 0.1 & 0.001 & Enzyme-ATP complex binds substrate\\
E$_2$-KKK-PP + ATP $\rightleftharpoons$ E$_2$-ATP-KKK-PP & 0.1 & 0.001 & Enzyme-substrate complex binds ATP\\
E$_2$-ATP-KKK-PP $\rightleftharpoons$ KKK-P + E$_2$-ADP + 2P$_i$ & 0.1 & 1e-30 & KKK-PP is dephosphorlyated\\
E$_2$-ADP $\rightleftharpoons$ E$_2$ + ADP & 0.1 & 1e-30 & Enzyme is regenerated\\
\end{longtable}
\end{center}

\newpage

\subsection*{Supplementary Figures}

\begin{figure}[h!]
\centering
\includegraphics[width=0.6\textwidth]{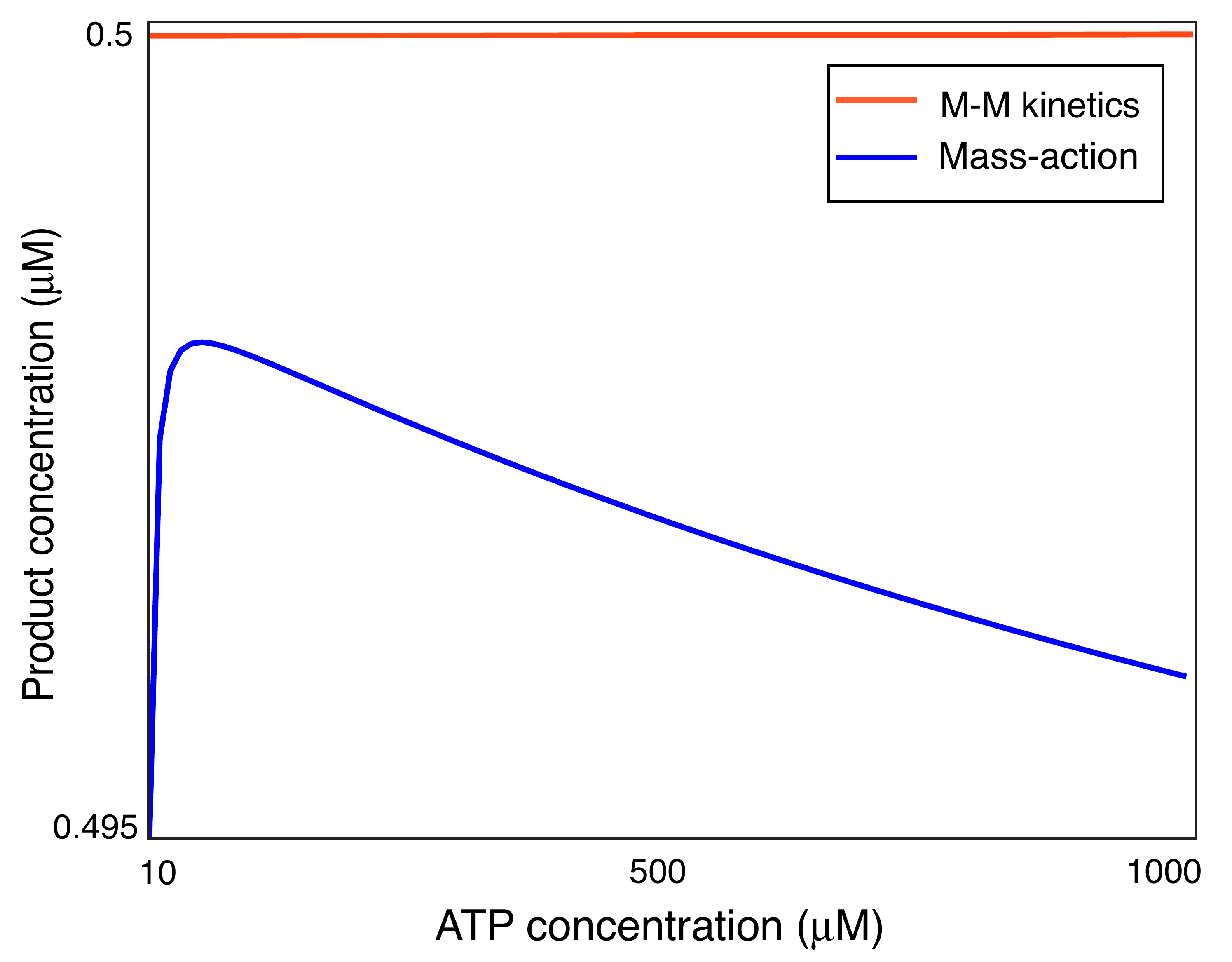}
\caption{Equilibrium product concentration in the two enzyme model (Figure \ref{fig:loops}A) at high ATP values. Red and blue lines show concentrations predicted by Michaelis-Menten kinetics and our model, respectively. As in other modules considered, a decline in equilibrium product concentration is observed in excess of a certain ``optimal'' ATP concentration (here, $\sim$ 70 $\mu$M). Note that values from the reversible mass-action model are always less than those predicted by Michaelis-Menten kinetics.}
\label{sfig:highATP2enz}
\end{figure}

\begin{figure}[h!]
\centering
\includegraphics[width=0.65\textwidth]{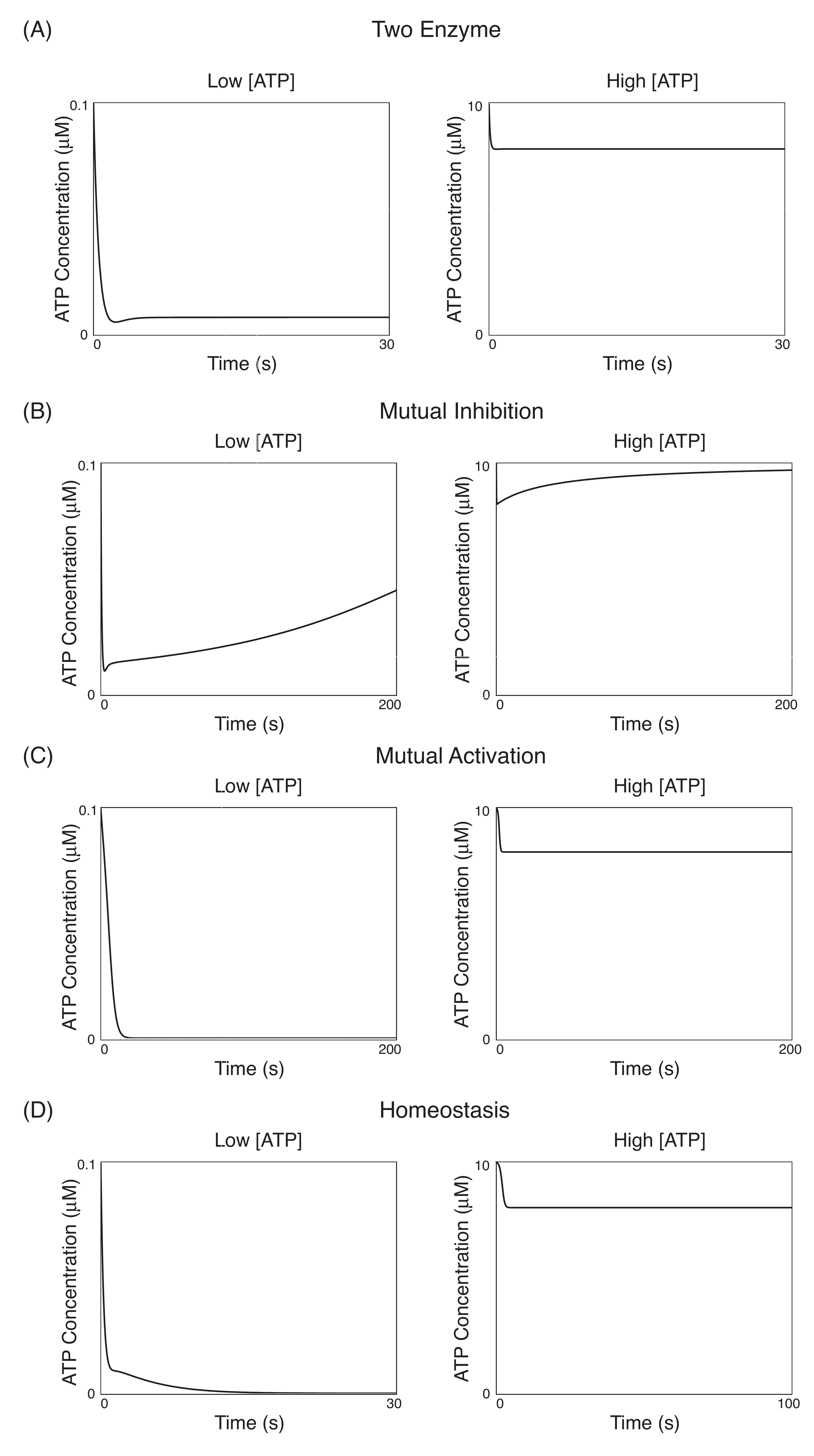}
\caption{Dynamics of ATP with low (left) and high (right) initial concentrations for modules shown in Figure \ref{fig:loops} in the main text. When initial concentrations are low, ATP is quickly depleted; when initial concentrations are high, a non-zero concentration is maintained at equilibrium. In the mutual inhibition module \textit{(B)}, accumulation of free phosphate from dephosphorylation allows for recycling of ATP.}
\label{sfig:atp}
\end{figure}

\begin{figure}[h!]
\centering
\includegraphics[width=0.65\textwidth]{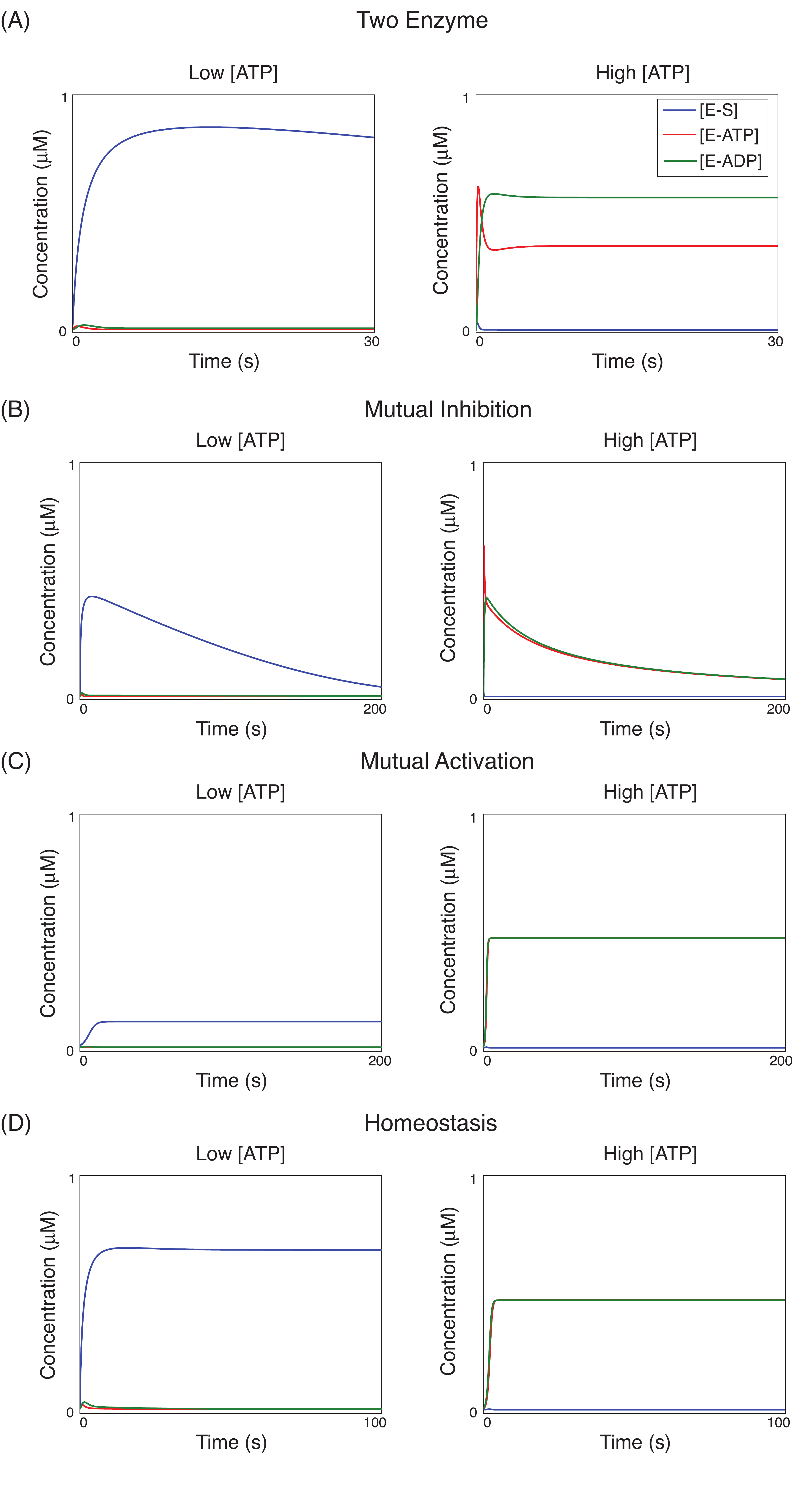}
\caption{Complex formation with low (left) and high (right) ATP concentrations for modules shown in Figure \ref{fig:loops} in the main text. Sequestration generally occurs in the form of E-S when ATP is low, and in the forms of E-ATP and E-ADP when ATP is high. In the mutual inhibition module (B), ATP cycling due to dephosphorylation allows for escape of free enzyme from complexes. Note that for all modules except homeostasis, the course of the two enzymes are symmetric. For this motif, results for E$_2$ are shown.}
\label{sfig:complex}
\end{figure}

\begin{figure}[h!]
\includegraphics[width=\textwidth]{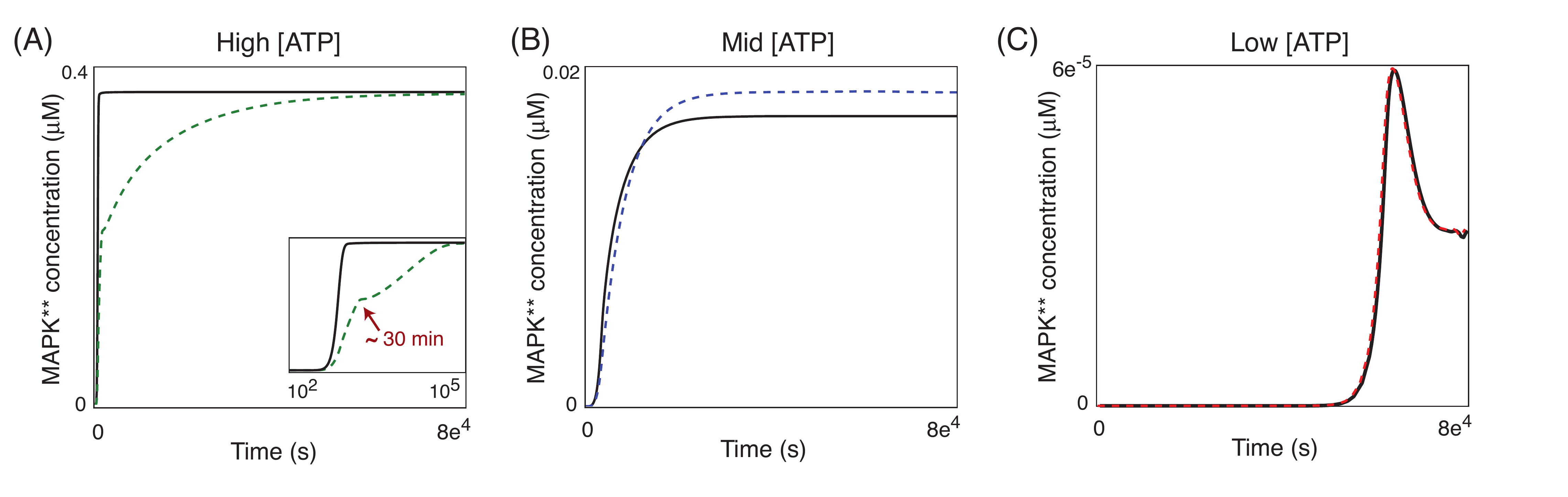}
\caption{ATP concentration affects the efficacy of feedback on the MAP kinase activation network. Solid black lines show time courses without feedback, and dashed colored lines show time courses when a feedback loop (shown in light gray in Figure \ref{fig:Mapk}) is activated \citep{bhalla1999emergent}. Time courses show the system response to a moderate stimulus (E$_1 = 0.001$ $\mu$M) in \textit{(A)} high, \textit{(B)} mid, and \textit{(C)} low ATP environments.}
\label{sfig:MAPKfdbk}
\end{figure}

\begin{figure}[h!]
\includegraphics[width=\textwidth]{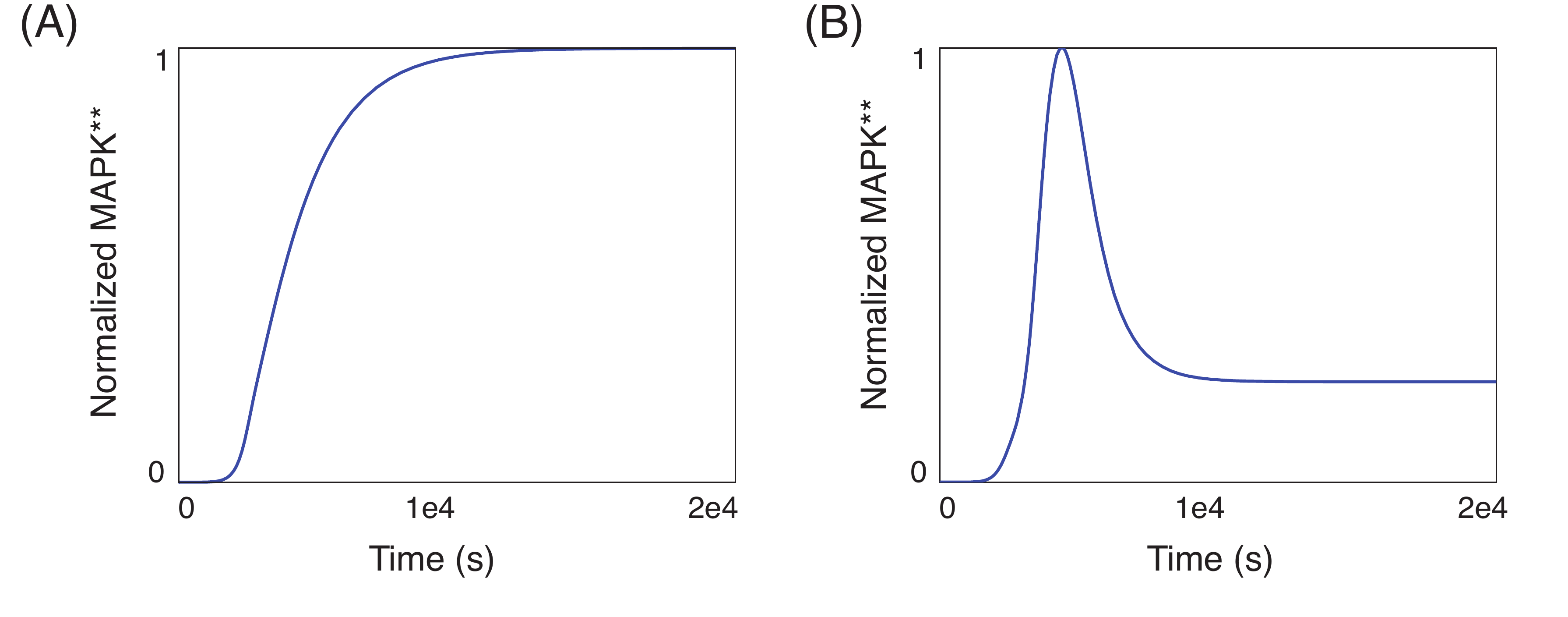}
\caption{Feedback, along with stimulus and ATP concentration, can affect the duration of MAP kinase activation. Responses without \textit{(A)} and with \textit{(B)} the feedback loop from \citep{bhalla1999emergent} are shown. Both responses are with [E$_1$]$|_{t=0}$ = 0.01 $\mu$M and [ATP]$|_{t=0}$ = 0.1 $\mu$M. Feedback loop was strengthened by multiplying $k_{\text{on}}$ rates by a factor of 100.}
\label{sfig:MAPKfdbk2}
\end{figure}

\end{document}